\documentclass[%
 reprint,
 twocolumn,
footinbib,
 amsmath,amssymb,
 aps,
 pra,
floatfix,
]{revtex4-2}

\usepackage{microtype}

\usepackage{algpseudocode}
\usepackage{algorithm}
\usepackage{lineno}
\usepackage{tabularx}
\usepackage{booktabs}
\usepackage{threeparttable} 
\usepackage{graphicx,
            caption,
            subcaption,
            ragged2e,
            xcolor,
            xspace,
            textcomp,
            hyperref,
            scrextend, 
            placeins, 
            tabularx,
            } 
\newcolumntype{M}[1]{>{\centering\arraybackslash}m{#1}} 

\newcommand{\mycolour}{black}

\usepackage[nospace]{varioref}
\usepackage[nolist]{acronym}
\usepackage[per-mode=symbol-or-fraction,output-decimal-marker={.}]{siunitx}
    \DeclareSIUnit\gauss{G}

\DeclareCaptionLabelSeparator{dot}{. }
\makeatletter
\def\justified{
	\let\\\@normalcr
	\@rightskip\z@skip \rightskip\@rightskip
	\leftskip\z@skip
	\parindent 0em\relax
	\setlength{\parfillskip}{0pt plus 1fil}}
\DeclareCaptionJustification{justified}{\justified}

\hypersetup{colorlinks=true, linkcolor=blue,citecolor=blue,urlcolor=blue}

\newcommand{\tx}{\text}
\newcommand{\e}{\tx e}

\newcommand{\ii}{\tx i}
\newcommand{\dd}{\tx d}

\usepackage{scalerel}
\usepackage{tikz}
\usetikzlibrary{svg.path}
\definecolor{orcidlogocol}{HTML}{A6CE39}
\tikzset{
  orcidlogo/.pic={
    \fill[orcidlogocol] svg{M256,128c0,70.7-57.3,128-128,128C57.3,256,0,198.7,0,128C0,57.3,57.3,0,128,0C198.7,0,256,57.3,256,128z};
    \fill[white] svg{M86.3,186.2H70.9V79.1h15.4v48.4V186.2z}
                 svg{M108.9,79.1h41.6c39.6,0,57,28.3,57,53.6c0,27.5-21.5,53.6-56.8,53.6h-41.8V79.1z M124.3,172.4h24.5c34.9,0,42.9-26.5,42.9-39.7c0-21.5-13.7-39.7-43.7-39.7h-23.7V172.4z}
                 svg{M88.7,56.8c0,5.5-4.5,10.1-10.1,10.1c-5.6,0-10.1-4.6-10.1-10.1c0-5.6,4.5-10.1,10.1-10.1C84.2,46.7,88.7,51.3,88.7,56.8z};
  }
}
\newcommand\orcidicon[1]{\href{https://orcid.org/#1}{\mbox{\scalerel*{
\begin{tikzpicture}[yscale=-1,transform shape]
\pic{orcidlogo};
\end{tikzpicture}
}{|}}}}

\usepackage[ngerman, main=british]{babel}        
\useshorthands{"}

\makeatletter
\@ifpackageloaded{babel}%
    {%
        \addto\extrasamerican{%
            \languageshorthands{ngerman}%
			}%
		\addto\extrasenglish{%
            \languageshorthands{ngerman}}
        \addto\extrasbritish{%
            \languageshorthands{ngerman}}%
    }{\relax}
\makeatother

\begin{document}


\author{Maximilian Sohmen~\orcidicon{0000-0002-5043-2413}}
\author{Maria Borozdova}
\author{Monika Ritsch-Marte~\orcidicon{0000-0002-5945-546X}}
\author{Alexander Jesacher~\orcidicon{0000-0003-4285-9406}}
    \email[Electronic address: ]{alexander.jesacher@i-med.ac.at}

\affiliation{Institute for Biomedical Physics, Medical University of Innsbruck, Müllerstraße~44, 6020~Innsbruck, Austria}

\title{Complex-valued scatter compensation in nonlinear microscopy}


\begin{abstract} 
Nonlinear, i.e., multi-photon microscopy is a powerful technique for imaging deep into biological tissues. 
Its penetration depth can be increased further using adaptive optics.
In this work, we present a fast, feedback-based adaptive-optics algorithm, termed C-DASH, for multi-photon imaging through multiply-scattering media. 
C-DASH utilises complex-valued light shaping (i.e., joint shaping of amplitude and phase), which offers several advantages over phase-only techniques: 
it converges faster, it delivers higher image quality enhancement, it shows a robust performance largely insensitive to the axial position of the correction plane, and it has a higher ability to cope with, on the one hand, scattering media that vary over time and, on the other hand, scattering media that are partially absorbing. 
Furthermore, our method is practically self-aligning.
We also present a simple way to implement C-DASH using a single reflection off a phase-only spatial light modulator.
We provide a thorough characterisation of our method, presenting results of numerical simulations as well as two-photon excited fluorescence imaging experiments. 
\end{abstract}

\date{\today}

\maketitle



\section{Introduction}

Imaging deep into biological tissue with cellular to subcellular resolution remains a challenge.
On its way to a deep-lying target structure, through the surrounding tissue, light gets scattered and wavefront errors accumulate. 
The development of multi-photon microscopes in the 1990s marked an important step towards deep-tissue imaging~\cite{denk1990two, hell1996three, sanderson_multiphoton_2023}, and the penetration depth has been further increased using adaptive optics (AO)~\cite{feierabend2004coherence, cha_shack-hartmann_2010, debarre_image-based_2009, jesacher_adaptive_2009, ji2010adaptive, booth2014adaptive, bueno_adaptive_2019, rodriguez2021adaptive, yao2023construction}.
For a long time, AO research has concentrated mainly on the regime of weak aberrations, where each photon is scattered less than once on average (i.e., the tissue thickness is less than the scattering mean free path, $l<l_s$).
Nowadays, AO research is increasingly directed at correcting turbid media in the multiple-scattering regime~\cite{tang2012superpenetration, papadopoulos2017scattering, yoon_deep_2020, berlage_deep_2021, may2021fast, qin_deep_2022, gigan2022roadmap, lim2022adaptive}. 
Such approaches are often referred to as `scatter compensation'.

In view of the the well-documented progress in optics research laboratories, why is scatter compensation not used in routine biomedical microscopy today?
Two persisting key challenges are, on the one hand, the low correction speed and, on the other hand, the small isoplanatic patch size, i.e., the area of an image for which the applied correction actually brings an improvement.
Concerning correction speed, in the multiple-scattering regime the lack of light of high spatial coherence generally prevents direct wavefront sensing using, e.g., a Shack-Hartmann sensor. 
Hence, one depends on indirect (=~feedback-based or sensorless) wavefront sensing~\cite{kubby2013adaptive}. 
In indirect wavefront sensing, a quality criterion [such as the intensity of multi-photon excited fluorescence (MPEF), or the image sharpness] is increased step by step through many sequential test measurements.
Due to the many optimisation steps, such iterative approaches are typically much slower than direct wavefront sensing. 
Concerning the struggle for larger isoplanatic patch sizes, besides technically challenging approaches such as multi-conjugate AO, one ansatz is to stitch many small patches (with individual wavefront corrections) together to obtain a larger, corrected image (see, e.g., Ref.\,\cite{may2021simultaneous}). 
This, however, makes the whole procedure even more time-consuming and illustrates the fundamental need for fast converging algorithms.

Amplitude aberrations are another phenomenon associated to strong optical disturbances, as is well known in astronomical AO~\cite{crepp2020measuring}. 
Amplitude aberrations arise even if the scattering medium is non-absorbing, simply by the propagation of uneven wavefronts. 
The twinkling of stars in the night sky is a manifestation of this effect many of us are familiar with. 
Intuitively, it seems obvious that correcting both -- the amplitude as well as the phase part of a scattered light field -- should deliver a higher performance than (even `perfect') phase shaping alone.
Yet, most scatter compensation techniques to-date have relied on phase-only corrections. 
Rare exceptions include works on field conjugation using a volume hologram~\cite{yaqoob2008optical, liu2015optical} and on imaging through multimode fibres~\cite{gomes2022near}. 
Of note, also by phase-only light shaping in more than one plane along the optical axis (as in multi-conjugate AO) it becomes possible to correct amplitude aberrations.
In the microscopy context, multi-conjugate AO configurations have been studied numerically~\cite{kam_modelling_2007, simmonds_modelling_2013, wu_numerical_2015} and, increasingly, also in experiments~\cite{hampson2020closed, furieri2023large}, but limited by technical difficulty only for low-order aberrations. 
Yet, the main motivation for building such systems has been not to correct amplitude aberrations, but to achieve larger isoplanatic patch sizes. 

In this work, we present the first feedback-based multi-photon AO scheme that jointly corrects the amplitude and phase part of a scattered light field.
As we show, joint amplitude and phase shaping is more powerful than phase-only techniques in four main aspects: 
(1) it converges faster;
(2) it can deliver a higher image quality enhancement; 
(3) it delivers a robust performance that is largely insensitive to the axial placement of the correction plane (pupil- vs sample conjugate AO); 
(4) it can deal with scatterers that are partially time-varying and/or absorbing, as is the case for most live biological tissues. 
We show that our algorithm singles out temporally stable and/or less-absorbing light paths in such cases.  
Furthermore, as we show, our method is practically self-aligning, greatly facilitating the experimental implementation.
Our development is based on Dynamic Adaptive Scatter Compensation Holography (DASH)~\cite{may2021fast, may2021simultaneous}, a scatter compensation method recently developed in our group. 
Since the procedure presented in this work can be understood as DASH extended to complex-valued light modulation, we refer to it as \emph{complex} or C-DASH. 
We further present a simple experimental implementation of C-DASH for two-photon fluorescence imaging, using a single reflection off a phase-only liquid-crystal spatial light modulator (SLM).

\section{C-DASH background}

\begin{figure}[tb]
\centering\includegraphics[width=\linewidth]{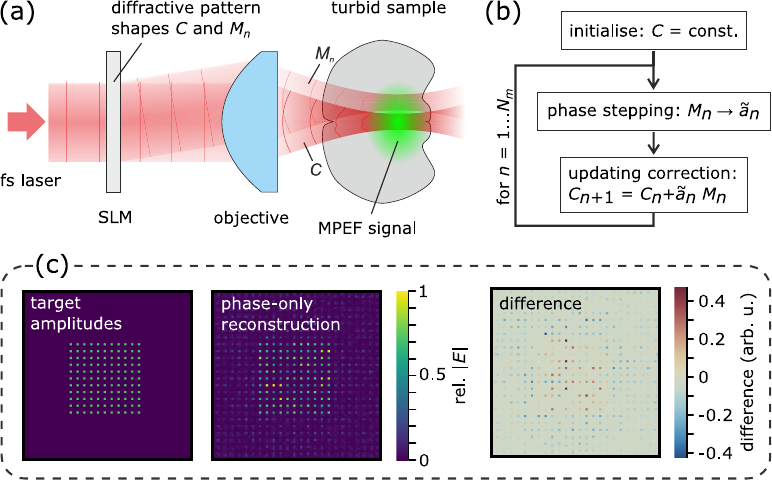}
\caption{\textbf{DASH/C-DASH principle and artefacts from phase-only light shaping.} 
(a)~A single diffractive phase mask displayed on an SLM shapes a corrected beam $C$ and a test mode $M_n$. The test mode is phase-stepped by updating the SLM.
(b)~Loop of C-DASH and DASH: Phase-stepping $M_n$ yields the complex-weighted power $\tilde a_n$, which allows to update the correction pattern $C_n$. 
The update of $C_n$ happens immediately after each mode measurement, up to a user-defined number $N_m$ of test modes.
(c)~Reconstructing a target amplitude distribution in the focal plane (\textit{left}) using phase-only light shaping in the Fourier plane gives rise to ghost spots and other inaccuracies (\textit{centre}). 
\textit{Right:} Difference between target and actual amplitude distribution.} 
\label{fig:DASH}
\end{figure}

Similar to other feedback-based AO schemes~\cite{vellekoop2008phase, conkey2012high, tang2012superpenetration, papadopoulos2017scattering, liu2017focusing, feldkhun2019focusing, qin_deep_2022}, in C-DASH a correction pattern is built up iteratively by analysing the interference between {\color{\mycolour} the reference beam (i.e., the wavefront with the current best correction, $C_n$) and a test mode, $M_n$, where $n \in \{1 \ldots N_m\}$ and $N_m$ is the number of test modes.}

In the case of C-DASH and DASH, the test modes are plane waves, see Fig.~\ref{fig:DASH}\,(a). 
{\color{\mycolour} Both methods use phase holography to shape $M_n$ and $C_n$, which offers the vital advantage that the power ratio $f$ between the test beam and the reference beam can be tuned freely.}
This is in contrast to methods using, e.g., a Hadamard basis ($f=1/2$) or the Continuous Sequential Algorithm (CSA, `pixel by pixel', $f=1/N_m$), where the power ratio $f$ is fixed.
The ability to tune $f$ ensures a sufficient signal-to-background ratio, which we have found to be crucial in real-world imaging applications~\cite{sohmen2022sensorless}.
{\color{\mycolour}As a result, our methods feature a particularly robust AO performance, as has been demonstrated using DASH for a range of artificial and biological samples~\cite{may2021fast, may2021simultaneous, sohmen2022sensorless}. 

In DASH, $C_n$ and $M_n$ are shaped holographically by disregarding the amplitude part and displaying only the phase part of the underlying complex fields on the SLM, which inevitably introduces wavefront inaccuracies [see, e.g., Fig.\,\ref{fig:convergence}\,(a)].}
In C-DASH, we avoid these wavefront inaccuracies by using joint phase and amplitude shaping.
For both C-DASH and DASH, we loop over the test modes $M_n$ in order of increasing spatial frequency. 
The phase-stepping procedure of a particular test mode $M_n$ yields a complex-weighted power $\tilde a_n$ (see Section~\ref{sec:FieldsReconstruction}), which allows to directly update the correction mask, $C_{n+1} = C_n + \tilde a_n M_n$. 
This basic procedure is illustrated in Fig.~\ref{fig:DASH}\,(b). 
A detailed description of the C-DASH algorithm can be found in the Methods.   

\subsection{Field reconstruction from phase stepping}
\label{sec:FieldsReconstruction}

As mentioned, C-DASH and DASH are indirect AO algorithms based on phase-stepping, which works as follows.
A reference beam $C_n$ and a test beam $M_n$ propagate through a scattering medium, interfere with each other and, if a fluorophore is hit, can generate MPEF.
One now measures the MPEF radiant power $P_\text{MPEF}(\phi_p)$ while varying the global phase 
$\phi_p = 2\pi p / N_p$ 
(where $p \in \{1 \ldots N_p\}$) between $C_n$ and $M_n$ somewhere \emph{before} the scattering medium, in our case using an SLM.
For the two-photon excited fluorescence (TPEF) results presented in this work, the order of nonlinearity is $\ell=2$ and, depending on the purpose, we chose values of $N_p$ between 3 and 24.

Let us assume the (static) reference beam and the (phase-stepped) test beam have amplitudes $u_\pm(\vec r)$, respectively, and interfere with each other inside a volume $V$.
We assume $\vec r = (x,y,z)^\top$, where the coordinate $z$ is along the optical axis.
Generally, $P_\text{MPEF}$ can contain signal contributions from all points inside $V$ where the fluorophore density $\rho(\vec r)$ is non-zero,
\begin{equation}
    P_\text{MPEF}(\phi_p) \propto \int_V \rho(\vec r) \,
    \left|u_+(\vec r) \, \e^{\ii\Delta \phi(\vec r)} + u_-(\vec r) \, \e^{\ii\phi_p}\right|^{2\ell} \dd^3 r.
    \label{eq:Pintegral}
\end{equation}
Here, $\Delta \phi(\vec r)$ is the relative phase between the two fields at the point of interference, $\vec r$, after propagation through the unknown medium, and we have omitted explicit time-dependence and neglected any kind of losses. 

For a moment, let us assume there is a single, brightest spot at $\vec r = \vec r_0$ within the fluorophore distribution $\rho(\vec r)$ that completely dominates the MPEF generation.
This assumption is often not too far from reality, since there is practically always an initial, brightest speckle, whose dominance will even grow during the AO procedure. 
In this case we can approximate
\begin{align}
    P_\text{MPEF}(\phi_p)
    &\propto \rho(\vec r_0) \,
    \left|u_+(\vec r_0) \, \e^{\ii\Delta \phi(\vec r_0)} + u_-(\vec r_0) \, \e^{\ii\phi_p}\right|^{2\ell} \nonumber \\
    &= \left|v_+ \e^{\ii\Delta \phi} + v_- \e^{\ii\phi_p}\right|^{2\ell}, \label{eq:DomSpeckApprox}
\end{align}
where we have included the fluorophore density $\rho$ in the amplitudes $v_\pm$ and dropped the explicit dependence on position.
Further, from the measurement of $P_\text{MPEF}(\phi_p)$ for the respective phase steps $\phi_p$ we can compute the mean power $\bar a$ and the complex-weighted power $\tilde a$ of the interfering excitation light,
\begin{align}
    \bar a &= \frac{1}{N_p}\sum_p P_\text{MPEF}^{1/\ell}(\phi_p) \qquad \text{and} \label{eq:MeanInt}\\
    \tilde a &= \frac{1}{N_p} \sum_p P_\text{MPEF}^{1/\ell}(\phi_p) \, \e^{\ii\phi_p}. \label{eq:ComplexInt}
\end{align}
Executing the sums under the dominant-speckle approximation (Eq.\,\ref{eq:DomSpeckApprox}) yields $\bar a = |v_+|^2 + |v_-|^2$ and $\tilde a = v_+^* v_-$, which lets us retrieve the amplitudes $v_\pm$ of the interfering excitation light fields at the dominating spot $\vec r_0$:
\begin{equation}
    v_\pm = \sqrt{\frac{1}{2} \left(\bar a \pm \sqrt{\bar a^2 - 4 |\tilde a|^2} \right)} 
\end{equation}
A fit of $P_\text{MPEF}(\phi_p)$ that reflects the dominant-speckle approximation is then given by
\begin{equation}
    P_\text{fit}(\phi_p) 
    = \left(\bar a + 2 \, |\tilde a| \cos(\Delta\phi-\phi_p) \right)^\ell,
    \label{eq:IFit}
\end{equation}
where $0 \leq 2\,|\tilde a| \leq \bar a$ by construction.
Note that $P_\text{fit}$ is fully characterised by three parameters: the mean excitation power $\bar a$, the modulation amplitude $2|\tilde a|$, and the relative phase given by
\begin{equation}
    \Delta \phi = \text{arg}(\tilde a). \label{eq:DelPhi} 
\end{equation}

Calculating $\bar a$, $\tilde a$, and $\Delta \phi$ via the analytic expressions (\ref{eq:MeanInt}, \ref{eq:ComplexInt}, \ref{eq:DelPhi}) is much faster and more stable than, e.g., performing an iterative least-squares fit to $P_\text{MPEF}(\phi_p)$.
Intuitively, the modulation amplitude $2\,|\tilde a|$ reflects the importance of the measured mode for MPEF generation at $\vec r_0$, while $\Delta\phi$ gives the relative phase where this mode contributes maximally.

For $N_p>3$ phase steps, the task of determining the three parameters $\bar a$, $\tilde a$ and $\Delta\phi$ is overdefined. 
In such a case, we can use the residuals of the fitted $P_\text{fit}(\phi_p)$ as an estimate of the noise on the measured $P_\text{MPEF}(\phi_p)$:
\begin{equation}
    \epsilon_\text{fit} = \sqrt{\frac{1}{N_p}\sum_p \left(P_\text{MPEF}(\phi_p) - P_\text{fit}(\phi_p) \right)^2}
    \label{eq:EpsilonFit}
\end{equation}
Of course, this definition of $\epsilon_\text{fit}$ is only strictly correct if the assumption of a single, dominating MPEF spot is fulfilled.
In practice, as mentioned, this is often the case, especially after the first tens of steps of our AO algorithms, when the brightest MPEF speckle is becoming more and more dominant.

When comparing the fit errors at different signal levels, the fit error can be normalised to MPEF power, giving a relative fit error:
\begin{equation}
    \epsilon_\text{rel} = \frac{\epsilon_\text{fit}}{\bar a^\ell}
    \label{eq:EpsilonRel}
\end{equation}

\subsection{Experimental implementation}
\label{sec:ExpImpl}

\begin{figure}[tb]
\centering\includegraphics[width=\linewidth]{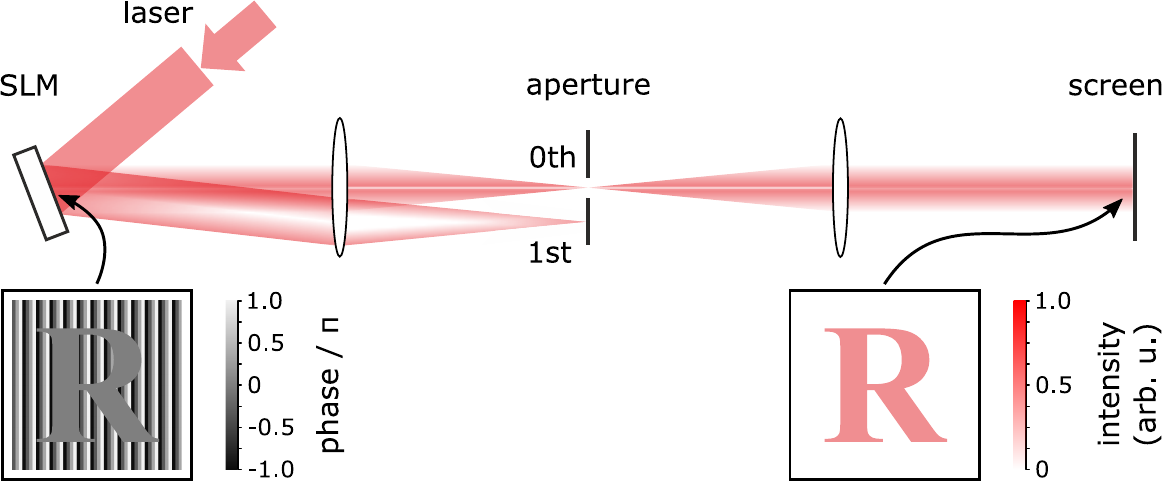}
\caption{\textbf{Complex-valued light shaping using a single, phase-only SLM.}
The target amplitude (in this example, a letter ``R'') is encoded in the modulation depth of a sawtooth grating displayed on the SLM.
Light reflected off a region with high grating amplitude is propagating off-axis and can be cut using an aperture after a lens.} 
\label{fig:complex_shaping_principle}
\end{figure}

\begin{figure*}[tb!]
\centering\includegraphics[width=\linewidth]{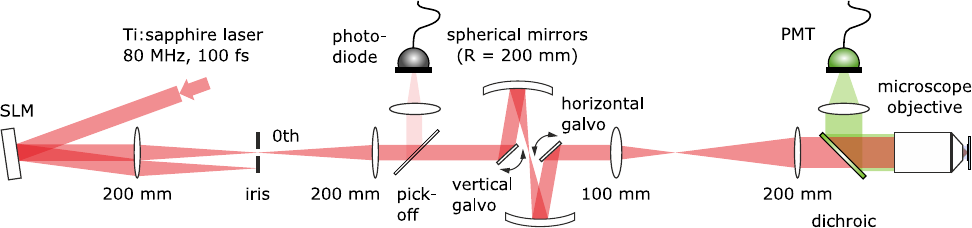}
\caption{\textbf{Experimental setup.} 
Sketch of the most essential parts in the excitation path of our TPEF microscope.
Fluorescence photons (green) are collected using the same objective as for excitation (red), split off using a dichroic mirror, and detected using a photo-multiplying tube (PMT).
} 
\label{fig:setup_suppl}
\end{figure*}

We realise joint amplitude and phase shaping using a single liquid-crystal SLM by imprinting our phase mask onto a sawtooth phase grating with locally varying modulation depth~\cite{kirk_phase-only_1971, bagnoud2004independent}. 
Recently, we have used this technique to boost the signal-to-background ratio for the Continuous Sequential Algorithm (CSA)~\cite{sohmen2022sensorless}. 
{\color{\mycolour} In the experiments presented here, we work with the light in the zeroth diffraction order of the SLM grating to avoid laser pulse dispersion.
For zero modulation depth (`flat' SLM), essentially all light goes into the zeroth order. 
For a sawtooth depth of $2\pi$, essentially all light is diffracted off the optical axis and dumped on an iris in the focal plane of a lens, as illustrated in Fig.~\ref{fig:complex_shaping_principle}. 
A locally varying sawtooth depth results in a locally varying intensity distribution inside the pupil.
For a given target intensity distribution, we compute the required local sawtooth grating depths using a clamp-and-adjust (CA) approach which is detailed in the Methods section. 
Let us mention here just three key facts: 
\textit{First}, by starting from an offset (non-zero) sawtooth grating depth (controlled via a parameter $\alpha$), the AO algorithm  can also increase (and not just decrease) the local intensity. 
\textit{Second}, CA returns a map of local sawtooth depths that produces a pupil intensity distribution which is (over a dynamic range that depends on $\alpha$) proportional to the target intensity distribution.
\textit{Third}, CA takes care to keep the total power coupled into the microscope objective constant during a whole C-DASH run by rescaling the sawtooth depth over the entire pupil.

A limitation of our single-reflection SLM approach is that inherently a fraction of the laser power is lost. 
A limitation of the CA procedure is its finite dynamic range (too high amplitudes are clipped). 
However, as we demonstrate, despite these limitations the SLM single-reflection approach allows to successfully perform AO using C-DASH.
Hence, these limitations need to be weighed against the experimental simplicity of our implementation.}
Further, let us highlight that using a more sophisticated experimental implementation, e.g., using two consecutive SLM reflections (off the same or two separate devices)~\cite{jesacher2008near, scholes2020lossless, barre_holographic_2022}, joint amplitude and phase shaping can be realised with minimal power losses. 

In the following we present results from numerical simulations and laboratory experiments. 
C-DASH and DASH experiments were performed on a home-built TPEF setup~\cite{may2021fast} using a $20\times$ water immersion objective (XLUMPLFLN20XW, Olympus Corp.). 
The wavelength of our femtosecond pulsed excitation laser (Mai Tai DeepSee\textsuperscript\textregistered, MKS Spectra-Physics) was set to $785$\,nm, which proved to work well for two-photon excitation of the samples presented in this work.
Our SLM (HSP1920, Meadowlark Optics, Inc.) was used in a pupil-conjugate configuration. 
Throughout each algorithm run, we monitored the optical power in the objective pupil using a power pick-off and a photodiode (see Fig.\,\ref{fig:setup_suppl}).
Fluorescence photons were collected through the same objective as for excitation. 
After collection, fluorescence photons were split off using a dichroic mirror, shortpass-filtered, and detected using a photo-multiplying tube (PMT, Hamamatsu Photonics K.K.).

\section{Static, non-absorbing scattering media}
\label{sec:Results}

Numerical simulations and experiments on static, non-absorbing scatter materials show that compared with phase-only light shaping, complex light shaping bears advantages in particular concerning three main factors:
\begin{samepage}
\begin{itemize}
    \item faster algorithm convergence,
    \item higher scatter compensation capabilities,
    \item robust performance largely independent of the axial position of the correction plane.
\end{itemize}
\end{samepage}
These results are laid out in the following Sections. 


\subsection{Simulation results: faster convergence}
\label{sec:FastConvergence}

\begin{figure*}[tb]
\centering\includegraphics[width=\linewidth]{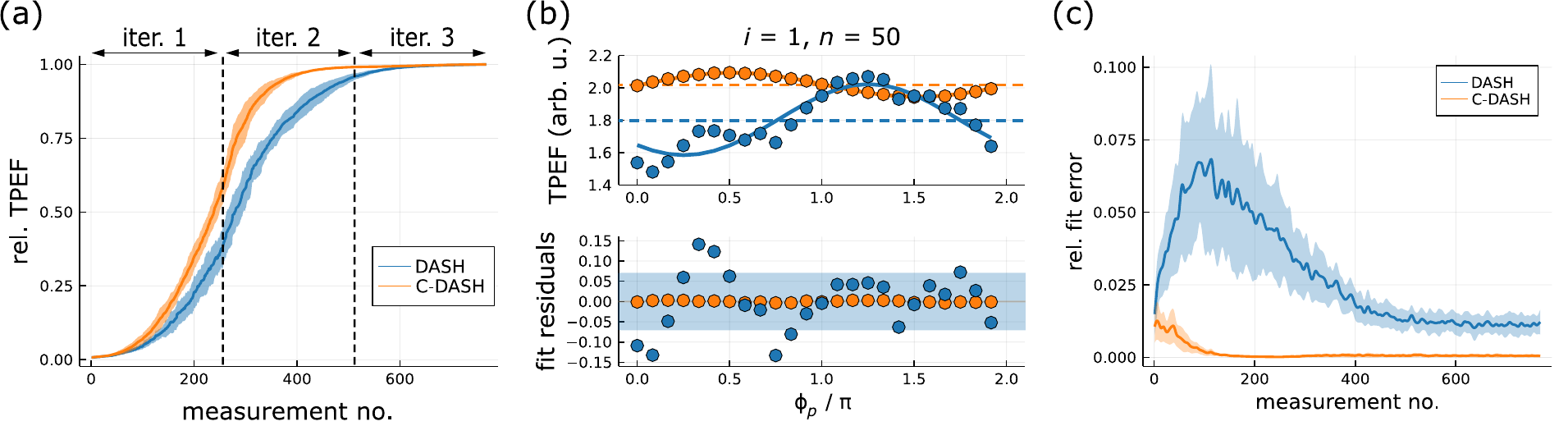}
\caption{\textbf{Convergence speed and light shaping accuracy.} 
(a)~TPEF signal for C-DASH (orange) and DASH (blue) for $N_i=3$ iterations of $N_m=256$~plane-wave modes. 
Solid lines give the mean, shaded bands the standard deviation over 5~runs with different random scatter masks (see main text). 
(b)~\textit{Top:} Phase stepping signal $P_\text{TPEF}$ (coloured circles), $\bar a^2$ (dashed horizontal lines) and $P_\text{fit}$ (solid lines) for a typical mode measurement (iteration~$i=1$, mode $n=50$), using a high number of phase steps, $N_p=24$.
\textit{Bottom:} Fit residuals (coloured circles) and $\pm\epsilon_\text{fit}$ (shaded bands). 
For C-DASH, the fit line and shaded band are mostly hidden by the plot markers.
(c)~Evolution of the relative fit error, $\epsilon_\text{rel} = \epsilon_\text{fit}/\bar a^2$, during the algorithm runs.
} 
\label{fig:convergence}
\end{figure*}

The difference in convergence speed between C-DASH (complex light shaping) and DASH (phase-only light shaping) is seen most easily in a simple numerical simulation as presented in Fig.\,\ref{fig:convergence}.
In this simulation (cf.~Refs~\cite{may2021fast, sohmen2022sensorless, sohmen2023_optofluidic}), the scatterer is modelled by a 2D plane of $16 \times 16 = 256$ pixels with uniformly random phases, located in a pupil-conjugate plane of a (square) objective lens.
The SLM (for correction) has the same size as the scatter mask and is also located in a pupil-conjugate plane. 
The sample is a 2D fluorescent layer in the focal plane. 
We test $N_m=256$ plane-wave modes in $N_i=3$ successive iterations.

Figure~\ref{fig:convergence}\,(a) shows the TPEF signal enhancement for C-DASH (orange) and DASH (blue), normalised to the signal from an aberration-free focus.
Both methods produce a steady, sigmoid-like increase of TPEF signal. 
We show later that the shape of the signal increase can be different; in the case of Fig.\,\ref{fig:convergence}\,(a), the well-behaved shape of the TPEF signal is related to the uniform randomness of the 2D scatter mask.
Further, both methods achieve full correction in our simulation, corresponding to a Strehl ratio of 1.0. 
This is expected, since the number of correction degrees of freedom matches the number of scattering degrees of freedom ($256$ each). 
However, we observe that C-DASH reaches 75\,\% of the total signal improvement already after about 280 mode measurements, whereas DASH reaches 75\,\% only after about 350 mode measurements. 
How relevant is this speed-up for applications?
Other, fast AO techniques~\cite{feldkhun2019focusing, liu2017focusing} require to detect the light transmitted through a scattering sample, with an abundance of photons available.
For C-DASH and DASH, in contrast, epifluorescence is sufficient, bearing clear advantages for application in microscopy.
However, compared with transmitted light, fluorescence photons are scarce. 
Hence, for feedback information from epifluorescence, longer dwell times are inevitable and it is particularly crucial to minimise the number of test measurements.
C-DASH does exactly this.
We attribute the faster convergence of C-DASH compared with DASH to its more precise light shaping capability, which leads to smaller errors in the mode measurements. 
However, before we interpret our findings, let us briefly recapitulate how a MPEF phase-stepping signal $P_\text{MPEF}(\phi_p)$ is expected to look like.

For linear fluorescence ($\ell=1$), the emitted power $P(\phi_p)$ has a sinusoidal shape (Eq.\,\ref{eq:IFit}), for a single as well as for multiple emitters, since the sum of multiple cosines with the same period (but different phases) is still a cosine with the same period (but different phase).
For non-linear fluorescence ($\ell\geq 2$), we have to distinguish the case of a singular emitter and the case of multiple emitters. 
For MPEF generated inside a single, localised spot, the $P_\text{TPEF}(\phi_p)$ will have the $\ell$-power cosine shape given in Eq.\,\ref{eq:IFit}.
If MPEF is generated in multiple spots, however, the shape can vary, since the sum of multiple cosines to the power of $\ell$ (with same period, but different phase) does not preserve the single-spot curve shape.
In an experiment, the measured shape of $P_\text{MPEF}(\phi_p)$ can deviate further from the ideal $\ell$-power cosine shape, e.g., due to measurement noise, or, as we show, as a product of inaccurate shaping of the reference and the test beam.

At the top of Fig.\,\ref{fig:convergence}\,(b), we plot the simulated phase-stepping signal $P_\text{TPEF}$ (coloured circles) from an exemplary mode measurement (iteration $i=1$, mode $n=50$).
A high number of phase steps ($N_p=24$) has been chosen to make deviations from the $\ell$-power cosine shape visible.
We see that the $P_\text{TPEF}$ oscillation produced by C-DASH is much closer to the expected shape ($P_\text{fit}$, solid lines) than DASH.
This is also mirrored in the fit residuals (coloured circles), plotted at the bottom of Fig.\,\ref{fig:convergence}\,(b), and the fit error, $\pm\epsilon_\text{fit}$ (Eq.\,\ref{eq:EpsilonFit}, shaded bands).

In Fig.\,\ref{fig:convergence}\,(c) we show the evolution of relative fit errors $\epsilon_\text{rel} = \epsilon_\text{fit}/\bar a^2$ (mean and standard deviation) over the 5 runs of C-DASH and DASH, respectively.
We observe a clear difference between C-DASH and DASH.
At the start of C-DASH, we find a finite $\epsilon_\text{rel}$, stemming from the fact that initially there are several strong speckles that contribute to $P_\text{TPEF}$ (i.e., the regime where $P_\text{fit}$ according to Eq.\,\ref{eq:IFit} is not really a proper model). 
However, during the first iteration of C-DASH, as the dominance of the strongest speckle grows, $\epsilon_\text{rel}$ quickly approaches zero and stays there. 
For DASH, $\epsilon_\text{rel}$ shows a different behaviour.
Starting from a similar initial level as for C-DASH, $\epsilon_\text{rel}$ grows substantially during the first iteration of DASH, before it slowly decreases again and settles at a finite level, yet by a factor of about $30$ higher than for C-DASH.
Since $\epsilon_\text{rel}$ is power-normalised, we can directly compare the relative fit errors for C-DASH and DASH.
As we do not see the intermediate increase for C-DASH, it seems 
likely that the intermediate increase and higher final level of $\epsilon_\text{rel}$ for DASH is due to an additional source of error:
the phase-only light shaping of DASH [$\Phi_\text{SLM} = \text{arg}(C)$] introduces wavefront inaccuracies, leading to `ghost spots' and other artefacts [Fig.\,\ref{fig:DASH}\,(c)] which affect the TPEF generation.
C-DASH ($\Phi_\text{SLM}$ according to Eq.\,\ref{eq:PhiSLM}, Methods), in contrast, does not suffer from such severe wavefront artefacts.

In summary, we observe that C-DASH generates a much cleaner phase stepping signal than DASH, which boosts the accuracy of the mode correction.
This is likely one of the main reasons why C-DASH converges faster than DASH.

In the simulation shown in Fig.\,\ref{fig:convergence}, the scatterers were such that both C-DASH and DASH were able to fully compensate them.
What happens if the scattering media are more complicated, such that they cannot be fully compensated anymore? 
This is investigated in the following Section.


\subsection{Simulation results: higher correction capability}
\label{sec:StrongerEnhancement}

What is to be expected about the correction capabilities of C-DASH compared with DASH? 
Generally, for a given number of corrected modes $N_m$, it is expected that complex-valued light shaping is superior to phase-only light shaping, since $2 N_m$ parameters (amplitude and phase) are optimised compared with just $N_m$ (phase only). 
However, the specific performance advantage will also depend crucially on the characteristics of the scatter medium as well as the position of the SLM within the optical path. 
For example, only in the case of a `thin' scatterer precisely conjugate to the SLM, it is (at least theoretically) possible to fully compensate the scatterer using a phase-only mask. 
If, conversely, the scatterer is `thick', there is generally no optical plane where a phase-only modulation can fully restore the beam quality. 
The particular thickness $l$ of a phase-only scatter medium above which amplitude corrections become important is determined by various factors, including the optical wavelength, the scatterer's spatial frequency content, and the scattering mean free path $l_s$ of a photon inside the scatter medium.

\subsubsection{Layout of the 3D numerical simulations}
\label{sec:Layout3Dsim}

To compare the performance between complex-valued and phase-only light shaping in a close-to-realistic two-photon imaging scenario, we performed 3D numerical simulations of DASH and C-DASH in Python. 
In these simulations, an excitation light cone (numerical aperture $\text{NA} = 0.5$) propagates through a non-absorbing 3D scatter medium.
Our scatter medium is modelled as a two-component conglomerate of materials with refractive indices $n_1$ and $n_2$, which form random-shaped 3D domains [see, e.g., the axial section in Fig.\,\ref{fig:simu_thick_graphs}\,(a)].
The average size of the refractive-index domains is on the order of $10\,\lambda_0$, where $\lambda_0$ is the exitation light vacuum wavelength.
Behind the scattering medium, in the (nominal) focal plane of the light cone, we place a 2D fluorescent layer.
Here, the excitation light generates a TPEF signal which we record over the algorithm runs.
We know from comparison to experiments that such a fluorophore-behind-scatterer model is representative of many biological specimens, such as fluorescence-labelled mouse brain \cite{may2021fast,sohmen2022sensorless}.
Greater detail on the simulations is provided in the Methods.

We simulate \emph{pupil-conjugate} as well as \emph{sample-conjugate} AO. 
In the pupil-conjugate case, the SLM is conjugate to the (circular-aperture) entrance pupil of the objective lens, i.e., Fourier-related to the fluorescent layer (and far away from the scatter medium). 
In the sample-conjugate case, the SLM is conjugate to a plane inside the scatter medium. 
We further study two different types of scatter media: a \emph{thin} scatterer of 4~$\lambda_0$ thickness and a \emph{thick} scatterer of 64~$\lambda_0$ thickness.

For non-absorbing scattering media that scatter predominantly into the forward direction, the average number of scattering events per photon, $\bar N_s = l/l_s$, can be calculated from the overlap integral $(\text{OI})$ between the undisturbed and the scattered field, $\bar N_s = -\ln(\text{OI})$~\cite{sohmen2022sensorless}.
Using this, we have chosen the refractive indices $(n_1, n_2)$ of the two components of our scatter media such that the thin and the thick scatterer lead to a similar $\bar N_s$ on the order of 3 to 4. 
This allows to directly compare the results for the thin and the thick scatterer.
Each scenario has been repeated 5~times with a different random scatterer, and every algorithm run comprised $N_i=3$ iterations of $N_m=256$ plane-wave modes.

\subsubsection{Thin scattering media}
\label{sec:ThinScatterer}

\begin{figure}[tb]
\centering\includegraphics[width=\linewidth]{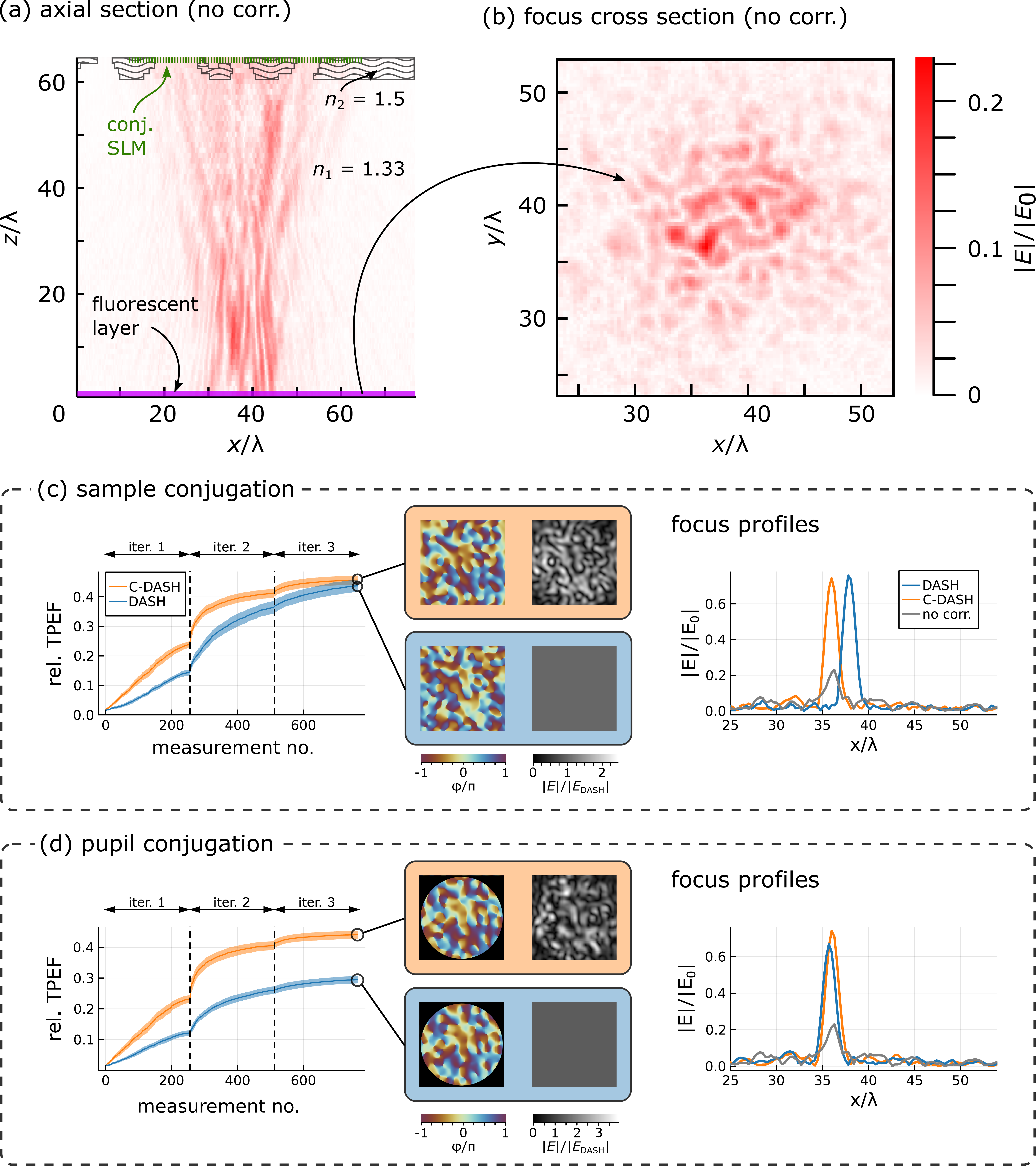}
\caption{\textbf{Simulation: compensating a thin scatterer.} 
(a) Axial section of a focusing light cone (red colour scale) propagating through a thin 3D scatter medium (wavy pattern), located $64 \, \lambda_0$ above the focal plane.
The scattering slice causes 3--4 scattering events per photon on average, decreasing the Strehl ratio to about $0.06$. 
(b) Cross section of the aberrated focus.
(c) Sample-conjugate AO [SLM imaged into the scattering layer at $z=64\lambda_0$; green dotted line in (a)]. 
\textit{Left:} C-DASH yields a 26-fold signal enhancement, DASH yields a 25-fold enhancement.
The $y$-axis is normalised to the TPEF of an aberration-free focus.
\textit{Centre:} Phase (red-blue) and amplitude part (black-white) of one run's final correction masks.
\textit{Right:} An example of the final focus profiles produced by C-DASH (orange) and DASH  (blue) as well as before correction (grey) for the scatterer of subfigure (a).
The amplitudes are normalised to an aberration-free focus ($E_0$).
(d) Pupil-conjugate AO [SLM far above plot window of (a)].
C-DASH yields a 25-fold signal enhancement, whereas DASH yields a 17-fold enhancement.
Note that our entrance pupil is circular, whereas for the sample-conjugate SLM mask we have allowed a square shape.
} 
\label{fig:simu_thin_scat}
\end{figure}

Figure~\ref{fig:simu_thin_scat} summarises the simulation results for a thin scatterer, located $60\,\lambda_0$ above the focal plane. 
Figure~\ref{fig:simu_thin_scat}\,(a) shows an axial section through the centre of the sample volume.
The scatter medium is located at the top.
The refractive indices of its two components are $n_1 = 1.33$ (white) and $n_2 = 1.50$ (grey wavy pattern).
The refractive index of the clear volume (white) between scatterer and focal plane is also $n_1=1.33$. 
This scatterer results in $\bar N_\text{scat} = 3.8(1.0)$ 
scatter events per photon (average and standard deviation over 5 different scatterers) and decreases the Strehl ratio to $0.06(0.02)$. 
The field amplitude of the (scattered) focusing light cone in the axial section is shown in white-red colour scale.
Figure~\ref{fig:simu_thin_scat}\,(b) shows the field amplitude in the focal plane.
For both field-amplitude plots, the colour bar is normalised to an aberration-free focus.

For \emph{sample-conjugate} AO (SLM imaged into the plane at $z=64 \,\lambda_0$), we adjust the side length of the SLM mask to the diameter of the (nominal) focusing cone at the respective $z$-position [see the green dotted line in Figure~\ref{fig:simu_thin_scat}\,(a)].
This ensures that the plane-wave test-mode phase varies by the correct amount over the (illuminated area of the) SLM mask (e.g., by $2\pi$ for the lowest mode after piston, $M_2$). 
Figure~\ref{fig:simu_thin_scat}\,(c) shows the TPEF signal evolution for DASH (blue) and C-DASH (orange), relative to the TPEF generated by an aberration-free focus. 
The lines and shaded bands represent the mean and standard error of the mean over five algorithm runs with different random scatterers.
Concerning the final signal level after $N_i=3$ iterations of $N_m=256$ modes, DASH and C-DASH perform roughly equally, as expected. 
For both methods, the TPEF is enhanced by a factor of about $25$ and the Strehl ratio increases from $0.06(0.02)$ to $0.65(0.04)$ for DASH and to $0.66(0.03)$ for C-DASH. 
The clear advantage of C-DASH is the fast early signal increase as discussed in the previous Section~\ref{sec:FastConvergence}.
Exemplarily, the plot in the centre gives the phase (red-blue) and the amplitude (black-white) of the final correction masks delivered by C-DASH and DASH for the particular scatterer shown in Fig.\,\ref{fig:simu_thin_scat}\,(a).
The plot on the right shows the focus profiles (cuts through amplitude maxima) before correction (grey) and after running C-DASH (orange) and DASH (blue) for the scatterer from Fig.\,\ref{fig:simu_thin_scat}\,(a).
We see that both DASH and C-DASH are able to restore a pronounced focus.
The two peaks seem similar in height for this particular run, but note that even a slightly higher field amplitude can result in a considerably higher TPEF signal due to the dependence to the power of 4.
Note also that the $x$-position of the peaks in the final focus profiles is different for C-DASH and for DASH; this shows that, to a certain degree, chance plays a role in the initial selection of which speckle becomes dominant. 
It seems likely that `ghost spots' and similar artefacts produced by the phase-only light shaping of DASH form an additional source of variability in this selection process.

Figure~\ref{fig:simu_thin_scat}\,(d) shows the corresponding TPEF signal evolution if we perform \emph{pupil-conjugate} AO for the same 5 thin scatterers as above.
Concerning the final TPEF level after $N_i=3$ iterations of $N_m=256$ modes, C-DASH performs significantly better than DASH. 
While pupil-conjugate C-DASH performs similar to the sample-conjugate case [25-fold TPEF enhancement, Strehl ratio $0.64(0.02)$], pupil-conjugate DASH delivers only a 17-fold TPEF enhancement and a Strehl ratio of $0.54(0.03)$.
The reason for this performance drop of DASH is that even a thin, non-absorbing (`phase-only') scatterer cannot be efficiently corrected by phase shaping alone in all cases where the SLM is not directly placed in a plane conjugate to the scatterer.  
Combined amplitude and phase shaping performs better in such cases.

As a last point, let us discuss the differences to the simulation with the 2D scatter plane from Section~\ref{sec:FastConvergence}. In the 3D simulations for Fig.\,\ref{fig:simu_thin_scat}, the scatterer was modelled on a grid of $256\times 256 \times 4$ points, whereas we used only $N_m=256$ plane-wave modes for correction. 
It is therefore unsurprising that our AO algorithms could not reach a Strehl ratio of~$1$ as before. 
Furthermore, let us qualitatively compare the shape  of the TPEF evolution curve for the pupil-conjugate runs [Fig.\,\ref{fig:convergence}\,(a) and Fig.\,\ref{fig:simu_thin_scat}\,(d)]. 
We observe that in contrast to the 2D scatterer, for the thin 3D scatterer the TPEF evolution deviates from the regular, sigmoidal shape. 
This is because the 3D scatterer does not add a uniformly random phase to each grid point, but has a spectrum of spatial frequencies in which low-frequency aberrations are more important than high-frequency aberrations.
This is in direct relation to the distribution of domain sizes in our simulated two-component scattering volumes.
Since we step through the plane-wave modes in order of increasing spatial frequency, the higher relevance of low-frequency aberrations leads to the step-like, accelerated increase with each start of a new iteration as visible in Fig.\,\ref{fig:simu_thin_scat}\,(d).
As we show, such a step-like increase is also found in experiments with various kinds of scatter media, where the distribution of sizes of the scattering structures often gives a higher weight to low spatial-frequency aberrations.


\subsubsection{Thick scattering media}
\label{sec:ThickScatterer}

\begin{figure}[tb]
\centering\includegraphics[width=\linewidth]{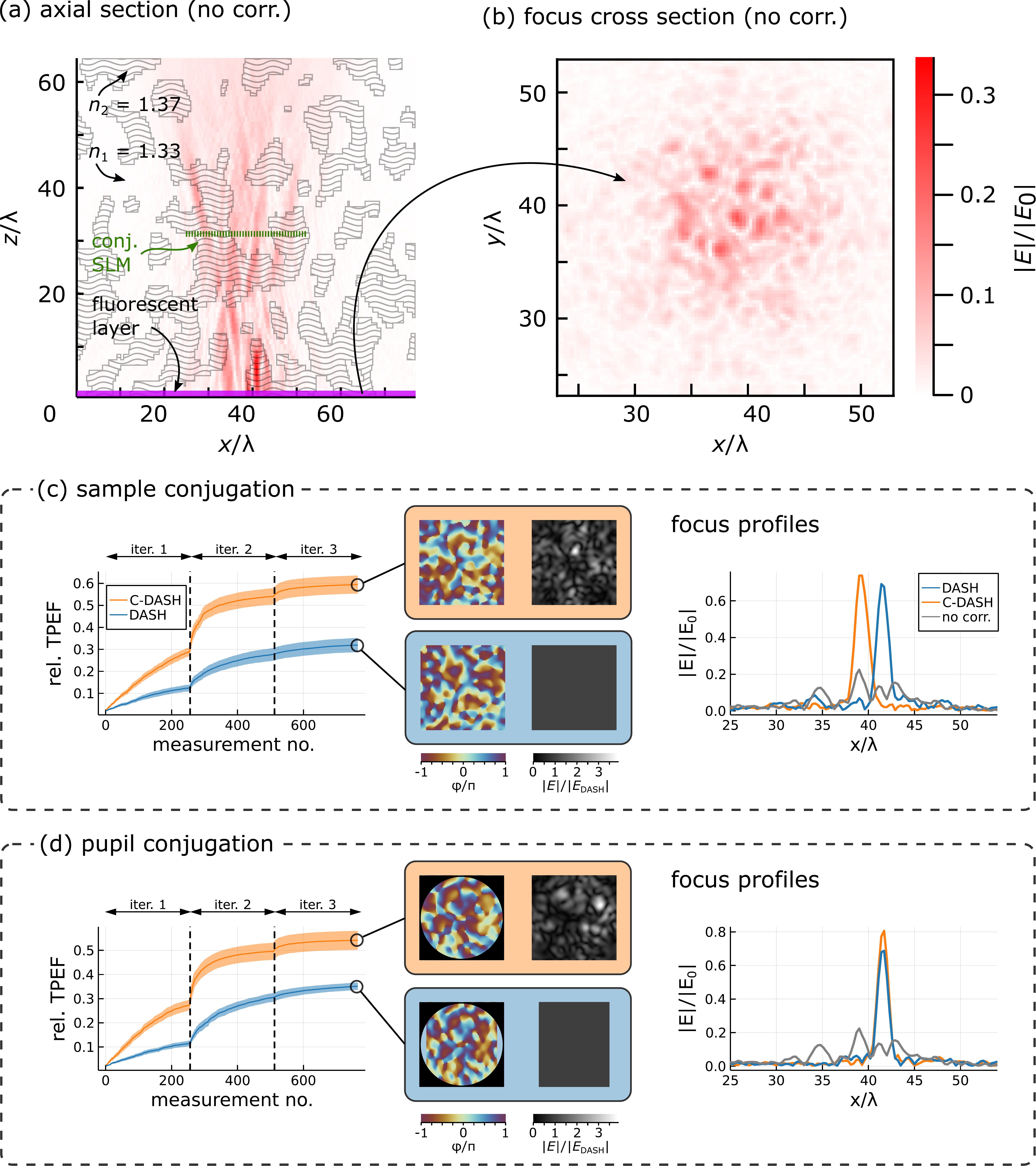}
\caption{\textbf{Simulation: compensating a thick scatterer.} 
(a) Axial section of a focusing light cone (red colour scale) propagating through a $64 \, \lambda_0$ thick 3D scatter medium (wavy pattern) above the focal plane.
The scatter medium causes 3--4 scattering events per photon on average, decreasing the Strehl ratio to about $0.08$. 
(b) Cross section of the aberrated focus.
(c) Sample-conjugate AO [SLM imaged into scatterer at $z=32\lambda_0$; green dotted line in (a)]. 
\textit{Left:} C-DASH yields a 24-fold TPEF signal enhancement, whereas DASH yields a 14-fold enhancement. 
The $y$-axis is normalised to the TPEF of an aberration-free focus.
\textit{Centre:} Phase (red-blue) and amplitude part (black-white) of one run's final correction masks.
\textit{Right:} An example of the final focus profiles produced by C-DASH (orange) and DASH (blue) as well as the profile before correction (grey) for the algorithm runs corresponding to subfigure (a).
The amplitudes are normalised to an aberration-free focus ($E_0$).
(d) Pupil-conjugate AO [SLM far above plot window of (a)]. 
C-DASH yields a 23-fold signal enhancement, whereas DASH yields a 15-fold enhancement.
} 
\label{fig:simu_thick_graphs}
\end{figure}

Figure~\ref{fig:simu_thick_graphs}\,(a) illustrates the situation for a light cone focusing into a $64 \, \lambda_0$ thick scatterer. 
We set the refractive indices of the two scatterer components to $n_1 = 1.33$ and $n_2 = 1.37$, respectively. 
This  results in $\bar N_s = 3.3(0.5)$ 
scatter events per photon and a Strehl ratio of $0.08(0.01)$ (average over 5 random scatterers), comparable to the thin scatterer with higher refractive index contrast (Section~\ref{sec:ThinScatterer} above).

In contrast to the thin scatterer, for the thick scatterer we do not observe stark differences between the performances for sample-conjugate and pupil-conjugate AO. 
For both cases, C-DASH achieves a much stronger signal enhancement than DASH.
Sample-conjugate C-DASH delivers a 24-fold TPEF enhancement, [Strehl ratio $0.80(0.01)$], whereas sample-conjugate DASH delivers only a 14-fold TPEF enhancement [Strehl ratio $0.62(0.06)$].
Pupil-conjugate C-DASH delivers a 23-fold TPEF enhancement [Strehl ratio $0.75(0.01)$], whereas pupil-conjugate DASH delivers only a 15-fold TPEF enhancement [Strehl ratio $0.62(0.04)$].
This points at a clear advantage of complex-valued over phase-only light shaping for thick scatterers. 

\medskip

Comparing the results for the thin and the thick scatterer, it appears that C-DASH is much more insensitive to factors like, on the one hand, the exact location of the SLM (sample- vs pupil conjugate AO) and, on the other hand, the thickness of the scattering medium. 
This can be intuitively understood as a direct consequence of the joint amplitude- and phase-shaping capability of C-DASH: 
As mentioned above, propagation of light alone is sufficient for turning phase-only aberrations into a mix of phase and amplitude differences. 
Therefore, in all cases where the scatterer is not just a thin slice with the SLM exactly conjugate to it, joint amplitude and phase shaping is superior to phase-only light shaping.

\subsection{Experiment: fluorescent microbeads}

\begin{figure}[tb]
\centering\includegraphics[width=\linewidth]{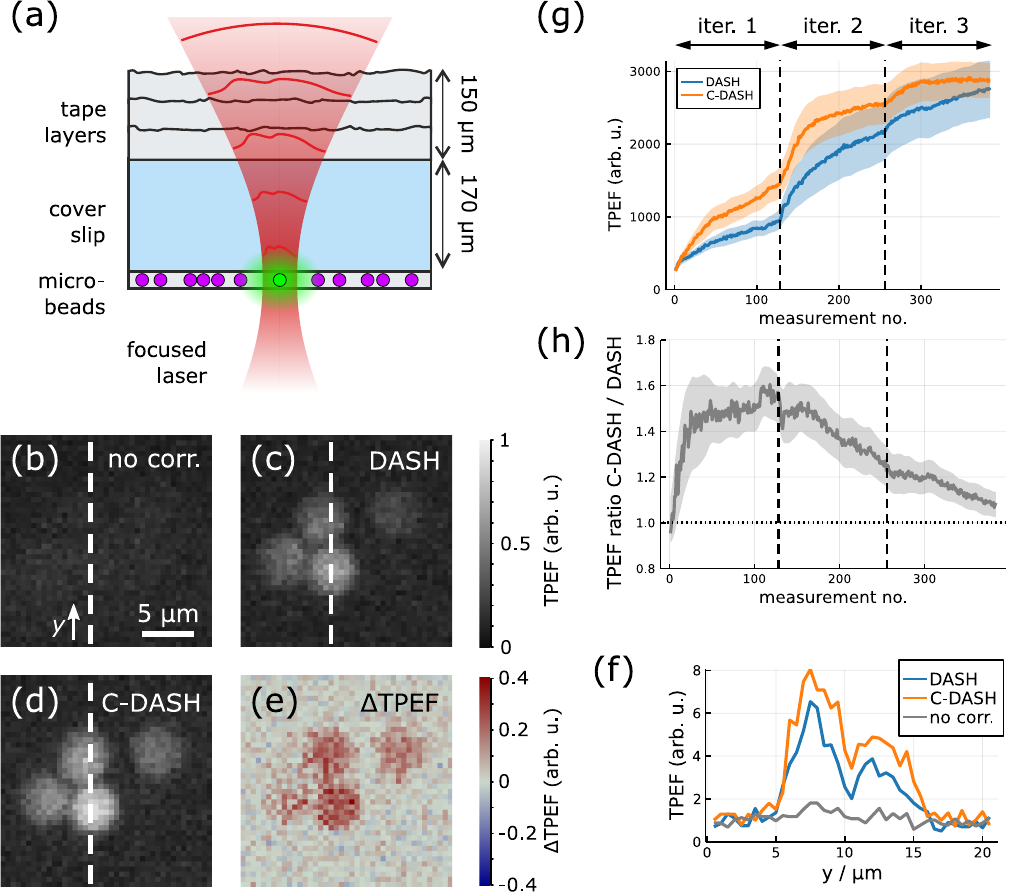}
\caption{\textbf{Experiment: fluorescent microbeads.} 
(a) Sketch of the sample. Fluorescent beads of 4\,µm diameter, embedded in liquid mounting medium, are covered by a glass slip and three layers of scattering adhesive tape. 
(b--d) Examples of TPEF scan images of the beads. 
(b) Without correction. (c) After 3 iterations of DASH. (d) After 3 iterations of C-DASH.
(e) The TPEF difference shows that the images (c) and (d) are properly centred.
(f) TPEF profiles along the white-dashed lines in subfigures (b--d).
(g) TPEF signal evolution while running DASH (blue) and C-DASH (orange). Lines give the mean, shaded bands the standard error of the mean over 6 independent runs with different target beads.
(h) Signal ratio between C-DASH and DASH. 
The grey line gives the mean, the shaded band the standard error of the mean for the same data as in subfigure (g).
} 
\label{fig:results_beads}
\end{figure}

For this experiment, we prepared a sample of fluorescent microbeads of 4\,µm diameter (TetraSpeck\textsuperscript\textregistered\ microspheres, Thermo Fisher Scientific, Inc.) immersed in a thin layer of liquid mounting medium (Thermo Fisher). 
As scattering medium we used three layers of matt adhesive tape, each 50\,µm thick, above the cover slip; see Fig.\,\ref{fig:results_beads}\,(a).
The tape layers strongly degrade the bead image, as shown in Fig.\,\ref{fig:results_beads}\,(b).

For the beads experiment, we set $\alpha=0.2$ (see Section\,\ref{sec:Methods_ComplexShaping}, Methods).
With this power overhead, the C-DASH algorithm can locally increase the intensity by a factor of up to 5.
Importantly, using our clamp-and-adjust procedure (see Methods), we make sure that the optical power reaching the objective pupil is constant over each C-DASH run, i.e., an increase in intensity in one region of the pupil is always balanced by a decrease in intensity in another region.
For direct comparability with C-DASH, we added a constant, uniform phase grating with $\alpha=0.2$ to all DASH phase masks throughout all DASH measurements. 
Remaining small variations of optical power in the objective pupil were recorded using a pick-off photo diode (see Fig.\,\ref{fig:setup_suppl}).
This allowed to both actively compensate the pupil optical power by adjustment of the laser output power and numerically post-correct the measured TPEF intensity with respect to excitation power.

Figures~\ref{fig:results_beads}\,(b), (c), and (d) show an example TPEF bead image before correction, after $N_i=3$ iterations of DASH, and after C-DASH, respectively, correcting $N_m=128$ modes per iteration. 
Compared with the uncorrected image, DASH already brings a clear TPEF signal improvement.
C-DASH, however, delivers an even higher TPEF improvement.
This is also reflected in the difference between the C-DASH and the DASH image, shown in Fig.\,\ref{fig:results_beads}\,(e), as well as in the image cuts along the white, dashed lines, plotted in Fig.\,\ref{fig:results_beads}\,(f).

Figure~\ref{fig:results_beads}\,(g) shows the evolution of TPEF signal while running C-DASH (orange) and DASH (blue).
Lines give the mean, shaded bands the standard error of the mean over 6 independent correction runs with different target beads (and, accordingly, a different part of the scatter tape in the light path). 
As before, we observe that C-DASH shows a faster increase of TPEF signal, particularly in the early phase.
Concerning the final TPEF signal after $N_i=3$ iterations of $N_m=128$ modes, C-DASH yields an 11-fold enhancement, DASH yields a 10-fold enhancement.
We also see these trends reflected in Fig.\,\ref{fig:results_beads}\,(h), where we plot the TPEF signal ratio between C-DASH and DASH. 
As we have mentioned before, the performance of C-DASH and DASH depends not only on the scatter medium, but also on the fluorophore characteristics. 
In this particular case of a spatially localised microbead, comparatively sensitive to bleaching, DASH is able to gradually (but not entirely) catch up with C-DASH toward the end of the third iteration.
As we show, this needs not be the case for other kinds of fluorophores.
A clear advantage of C-DASH over DASH is the  fast signal increase in the beginning, most prominent at the end of the first iteration, by a factor of about 1.6.


\subsection{Experiment: dye layer}

\begin{figure}[tb]
\centering\includegraphics[width=\linewidth]{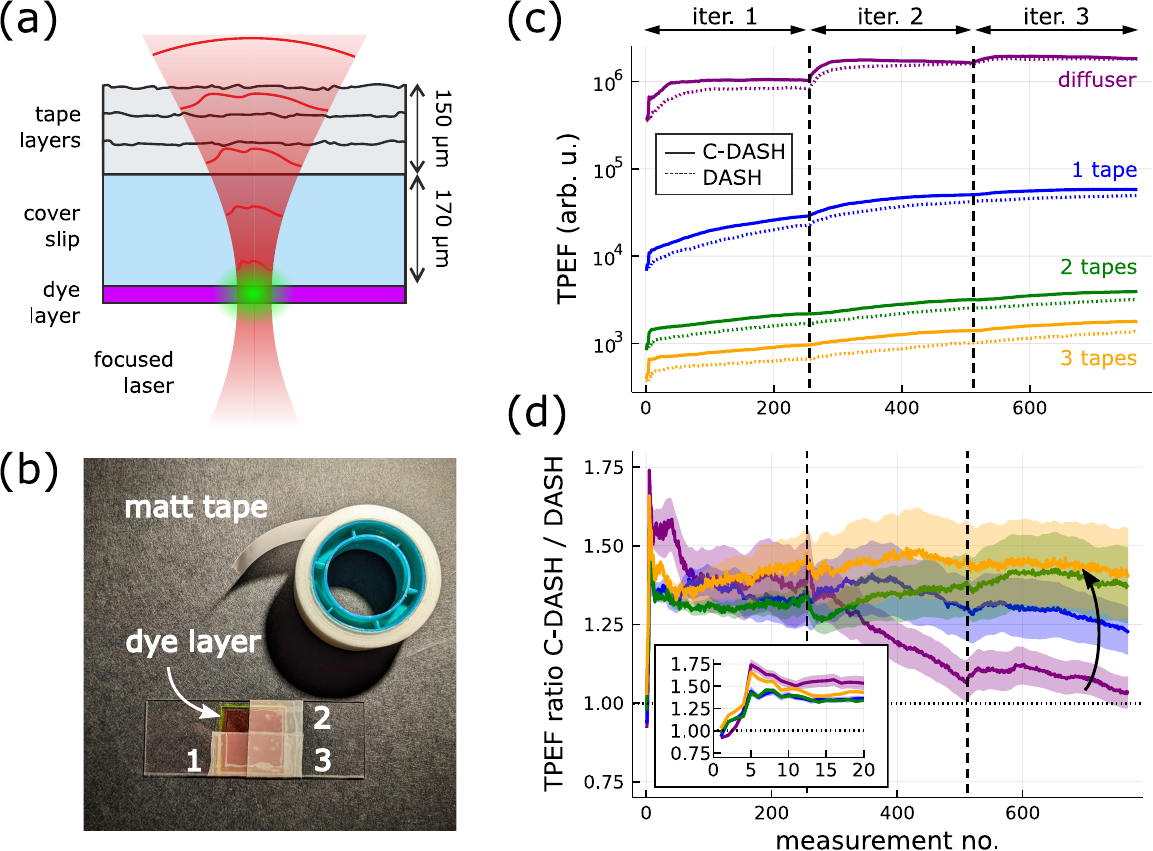}
\caption{\textbf{Experiment: dye layer.} 
(a) Sketch of one type of sample: A layer of fluorescent dye (rhodamine 6G) is covered by a glass coverslip and three layers of scattering adhesive tape. 
(b) Photograph of the matt tape and a typical sample. 
The reddish dye layer below the cover slip as well as the number of tape layers (1--3) are indicated. 
(c) Results for C-DASH and DASH corrections for 4 different scatterers: a gentle polymer diffuser (0.5$^\circ$ cone angle, Newport Corp.) and 1 to 3 layers of scattering adhesive tape, respectively.
Depending on the severity of the scatterer, the amount of generated TPEF varies over more than 3 orders of magnitude (note the logarithmic $y$-scale).
The different scattering scenarios are grouped by colour; solid lines correspond to C-DASH, dashed lines to phase-only DASH.
(d) The ratio between the TPEF signals for C-DASH and DASH [mean and standard deviation; same colours as in (c)]. 
C-DASH shows a faster initial enhancement than DASH. 
For gentle, thinner scatterers DASH reaches a similar final enhancement as C-DASH, for thicker scatterers it does not (black arrow).
The inset shows that the two algorithms always start from the same initial signal level.} 
\label{fig:tape_exp}
\end{figure}

Our dye layer experiment was conducted in a manner very similar to the beads experiment. 
A thin layer of Rhodamine~6G below a glass cover slip was used as fluorescent sample. 
We studied the performance of pupil-conjugate C-DASH and DASH for different levels of scattering severity. 
In order of increasing severity, our scattering media were either a gentle polymer diffuser (0.5$^\circ$ cone angle, Newport Corp.), placed in a plane conjugate to about $50$\,µm above the focal plane, or 1--3 layers of matt adhesive tape directly on top of the cover slip [see Fig.\,\ref{fig:tape_exp}\,(a,\,b)].

Figure~\ref{fig:tape_exp}\,(c) gives an overview of the algorithm performances for the different scenarios.
We plot the evolution of the TPEF signal as the algorithms proceed, where the scattering scenarios are grouped by colour; solid lines correspond to C-DASH (mean over 15 repetitions, each using a different region of the scatterer of the respective severity level), dotted lines to DASH (mean over 15 repetitions with the same scatterer regions as C-DASH).
Importantly, the TPEF signals plotted in Fig.\,\ref{fig:tape_exp}\,(c) are corrected for excitation power (i.e., we normalise to the square of the excitation power measured using our power pick-off).\footnote{NB that throughout this work, all plots of TPEF over algorithm runs are corrected for excitation power in the same way as here. 
We stress it at this point simply because it is of special importance when comparing different scattering severity levels in a single plot.}
This allows to combine all scattering scenarios into a single, semi-logarithmic plot and directly compare them. 
As we see, with the polymer diffuser (purple traces) the most TPEF per excitation power-squared is generated, i.e., the polymer diffuser causes the least scattering in our comparison. 
Note that the diffuser traces are also the most step-like, which is expected, as compared with the more violent scatterers, it has the least content of high spatial-frequency aberrations. 
Comparing C-DASH and DASH for the diffuser, we observe that the trace for C-DASH increases faster than for DASH, and only after $N_i=3$ iterations DASH has gradually (but not quite) caught up with C-DASH, where they reach a TPEF enhancement compared with the start value of a factor of about 5. 
Turning to the traces for 1--3 layers of scattering tape (blue, green, yellow, respectively), we observe that the general level of TPEF per excitation power-squared goes down by more than 3 orders of magnitude, indicating that the scattering becomes substantially more severe.
We also observe that with increasing level of scattering severity, the traces become less and less step-like, as expected when aberrations over the full range of spatial frequencies become important.
Concerning the tape samples, we observe again that C-DASH rises faster and higher (TPEF increase by factors 5 to 8 compared with start values) than DASH (factors 3 to 7).
For easier comparison between the respective performances of C-DASH and DASH, consider Fig.\,\ref{fig:tape_exp}\,(d), where we plot the ratios of power-corrected TPEF signal between C-DASH and DASH during the algorithm runs for the four scattering severity levels. 
For all scattering levels, we find that this ratio, starting from unity, initially quickly rises to values between $1.35$ and $1.75$ (see also the inset), confirming our previous observations that C-DASH tends to deliver a faster TPEF increase than DASH.
Compared with the diffuser case (purple), where DASH shows a tendency to catch up with C-DASH until the end of the third iteration, for increasing levels of scattering severity (blue, green, yellow), DASH increasingly struggles to catch up [black arrow in Fig.\,\ref{fig:tape_exp}\,(d)]. 
This is likely related to our observation from the 3D numerical simulations above that the thicker a scatter medium, the more crucial it becomes to correct amplitude aberrations.

\section{Temporally varying and/or absorbing scatterers}

Up to this point, we have restricted our discussion to scattering media which are both non-absorbing and constant in time.
In live tissue, however, these two conditions are often violated, which presents a big challenge for AO and imaging attempts.
Absorbing regions in the excitation light path, on the one hand, will not only reduce the effective power available for fluorophore excitation: if, locally, the absorbed power becomes too high, the sample might be burnt and destroyed.
A temporally varying scatterer, on the other hand, will make AO correction patterns invalid after a certain time. 
This can be particularly problematic for feedback-based AO due to the required algorithm run time: if the run time is longer than a tissue's timescale of decorrelation, correction patterns become invalid faster than they can be retrieved.
Nevertheless, research has shown that surprisingly long-lived correction patterns can be found~\cite{blochet2019enhanced}. 
The proposed explanation for this observed longevity is that in typical biological tissues there is more than a single decorrelation timescale.
While some regions of a sample, like pulsating blood vessels, will fluctuate on a short timescale, other regions might be comparatively stable.
AO algorithms can be designed to preferentially take into account scattering modes that are more stable over time and disregard scattering modes that fluctuate strongly {\color{\mycolour} (compare, e.g., Ref.\,\cite{mididoddi2023})}.

In this Section we show that joint phase and amplitude shaping boosts the algorithm's ability to bypass problematic regions inside the scattering medium, i.e., regions of strong temporal decorrelation and/or absorption. 
We present numerical simulation results for a medium where in one region the refractive index fluctuates temporally, and complement this with laboratory results from experiments with a partly-absorptive scatterer.

\subsection{Simulation: fluctuating refractive index}
\label{sec:SimFluctRI}

\begin{figure}[tb]
\centering\includegraphics[width=\linewidth]{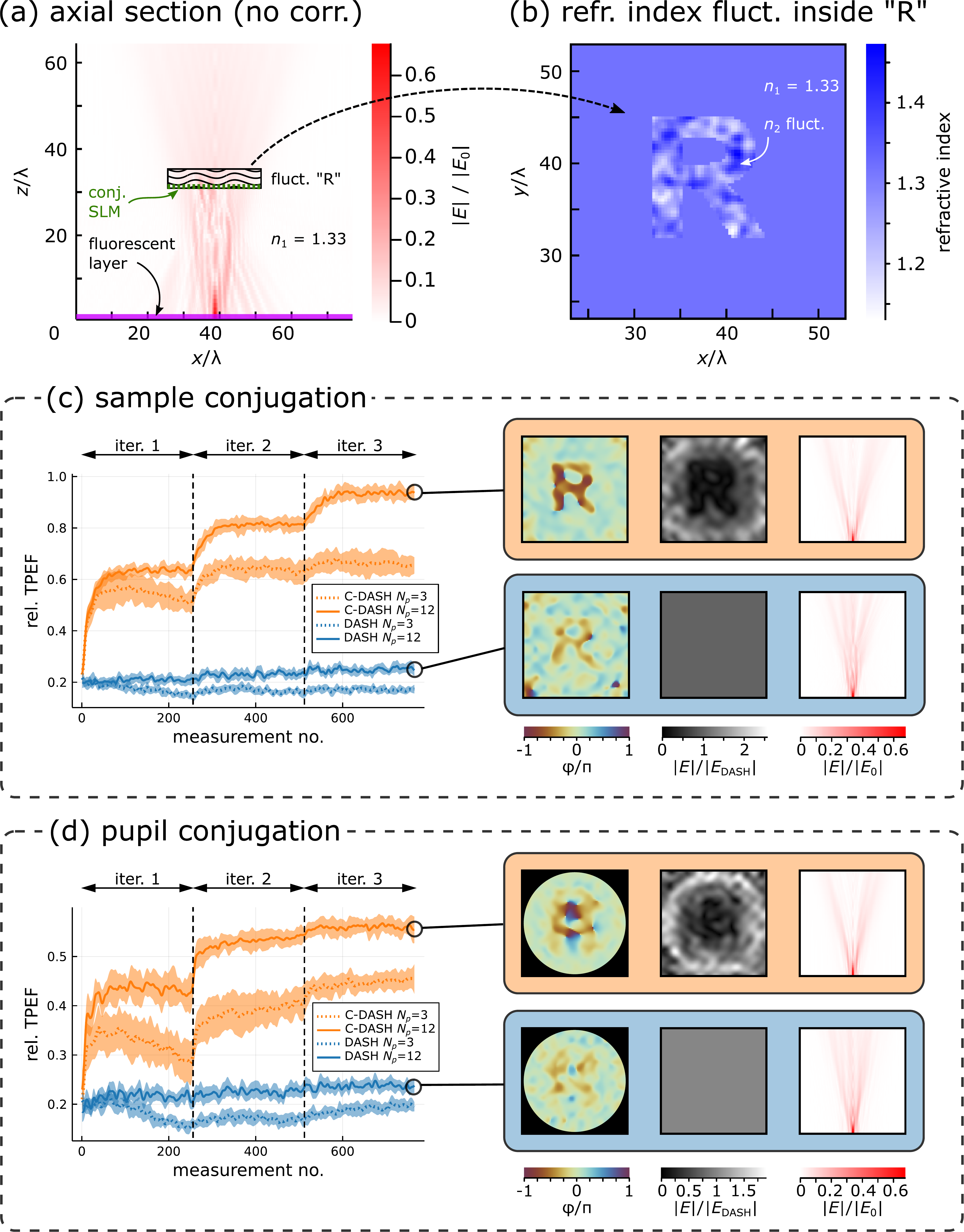}
\caption{\textbf{Simulation: compensating temporally varying scattering media.} 
(a)~Axial section (snapshot): Light is focused through a 3D volume (wavy-patterned box) of $6 \, \lambda_0$ thickness.
$|E|$ (white-red) is normalised to an aberration-free focus, $|E_0|$.
(b)~Cross-section of wavy-patterned box: 
Inside the ``R'', the refractive index distribution $n_2(x,y)_R$ fluctuates randomly from shot to shot (see main text).
(c)~Sample-conjugate AO [SLM = green dotted line in (a)]. 
\textit{Left:} 
TPEF signal evolution during C-DASH (orange) and DASH (blue); mean and standard deviation over 3~runs each. 
Dotted (solid) lines correspond to runs with $N_p = 3$ ($N_p = 12$).
\textit{Right:} Exemplary results for one run (each) with $N_p=12$.
Phase (red-blue) and amplitude part (black-white) of the final correction masks, as well as the light cone after AO.
(d)~Corresponding results for pupil-conjugate AO [SLM far above the plot window of (a)].} 
\label{fig:fluctuating_box}
\end{figure}

The setting for our numerical simulations is illustrated in Fig.\,\ref{fig:fluctuating_box}\,(a). 
A light cone is focused into a medium of refractive index $n_1 = 1.33$.
A volume in shape of a letter ``R'' and of $6\,\lambda_0$ thickness is placed centrally in this light cone. 
Inside the ``R'', the refractive index $n_2(x,y)_R$ varies over distances on the order of $2.5\lambda_0$ and spans values within $n_1 \pm 0.2$.
A new $n_2(x,y)_R$ is random-generated for each phase step. 
This mimics a fluctuation of refractive index inside the ``R'' on a timescale much shorter than the measurement of one test mode.  
We correct $N_m=256$ plane-wave modes in $N_i = 3$ iterations.

Let us first consider the sample-conjugate configuration, with the SLM plane imaged into the scattering volume at $z=32\lambda_0$ [green dotted line in Fig.\,\ref{fig:fluctuating_box}\,(a)].
On the left of Fig.\,\ref{fig:fluctuating_box}\,(c), we show the evolution of the TPEF signal during the algorithm runs of C-DASH (orange) and DASH (blue), relative to an aberration-free focus.
Dotted and solid lines correspond to runs with $N_p=3$ and $N_p=12$ phase steps, respectively. 
Lines give the mean, shaded bands the standard deviation over 3~algorithm runs each.
The respective signal enhancements are summarised in Table~\ref{tab:R_enhancements}.
In short, we see that C-DASH increases the TPEF up to more than 90\,\% of the aberration-free focus, whereas DASH delivers hardly any improvement. 
Additionally, the runs with $N_p=12$ deliver a better performance than the runs with $N_p=3$.
On the right, we show the phase (red-blue) and amplitude (black-white) part of the final compensation masks produced by C-DASH (top) and DASH (bottom), as well as axial cuts of the respective light cones after AO compensation (examples shown for the first out of 3 algorithm runs with $N_p=12$).
Here we observe that a dark ``R'' has appeared in the amplitude part of the C-DASH mask, i.e., C-DASH has taken light power away from the problematic, fluctuating region, and channelled it through the non-fluctuating regions.
This is also visible in the axial cut of the AO-corrected light cone.
DASH, of course, does not feature such amplitude shaping capabilities.

\begin{table}[tb]
    \caption{\textbf{Signal enhancements for a temporally varying scatterer -- fluctuating ``R''.}
    Mean and standard deviation of TPEF\textsubscript{final}$/$TPEF\textsubscript{initial} over 3 algorithm runs, for $N_p=3$ ($N_p=12$; in parentheses) phase steps.
    $N_i=3$, $N_m=256$; numerical simulation, see main text. 
    }
    \centering
    \begin{tabularx}{\columnwidth}{m{15mm} X X}
        \toprule
        \textbf{Conj.} & \textbf{C-DASH} & \textbf{DASH}\\
        \midrule
        pupil & $2.0\pm 0.3$ \quad ($2.9 \pm 0.3$) & $0.9\pm 0.1$ \quad ($1.2\pm 0.2$)\\
        sample & $3.4 \pm 0.4$ \quad ($4.5\pm 0.5$) & $0.7\pm 0.1$ \quad ($1.3\pm 0.2$)\\
        \bottomrule
    \end{tabularx}
    \label{tab:R_enhancements}
\end{table}

Figure~\ref{fig:fluctuating_box}\,(d) shows the corresponding results for pupil-conjugate AO [SLM far above plot window of Fig.\,\ref{fig:fluctuating_box}\,(a)].
In this case, C-DASH reaches about 55\,\% of the aberration-free TPEF, otherwise our observations are largely consistent with the sample-conjugate case.
Note also that we provide video animations of the fluctuating scatterer's effect on the light cone, before and after AO correction, in our Supplementary Material~\footnote{See Supplemental Material at [URL will be inserted by publisher] for the animated GIF videos and a corresponding description.}.

From the results shown in Fig.\,\ref{fig:fluctuating_box} we can draw two main conclusions. 
First, C-DASH considerably improves the TPEF signal for the temporally varying scatterer by guiding the light around the fluctuating region, whereas DASH largely fails to do so.
Second, a higher number $N_p$ of phase steps (solid vs dotted lines) speeds up the TPEF increase for C-DASH.
How can we understand this? 
Consider a specific scattering mode $M_n$ that fluctuates strongly (phase shifts $\gtrsim 2\pi$) on a timescale much shorter than the time needed to record $N_p$ phase steps.
Then the corresponding phase-stepping TPEF signal modulation 
($\propto |\tilde a|$,  Eq.\,\ref{eq:IFit}) 
will tend to zero and the mode's contribution ($\tilde a_n M_n$) to the sum of correction modes will vanish.
A similar argument holds for modes fluctuating on a timescale shorter than the time to complete one out of $N_i$ iterations of $N_m$ mode measurements.
Therefore, the higher the number of phase steps $N_p$ compared with the minimum number of 3, and the higher the number $N_i$ of algorithm iterations, the less the contribution of such modes to the final correction pattern.
As a consequence, the final correction pattern will predominantly contain scattering modes that are constant at least over the timescales defined by one mode measurement and one iteration of $N_m$ mode measurements, respectively. 

Note that analogous reasoning suggests that C-DASH will tend to disregard modes that are attenuated by absorption inside the scattering medium (following Section).

\subsection{Experiment: partially absorptive scatterer}
\label{sec:ExpAbsorptive}

\begin{figure}[tb]
\centering\includegraphics[width=\columnwidth]{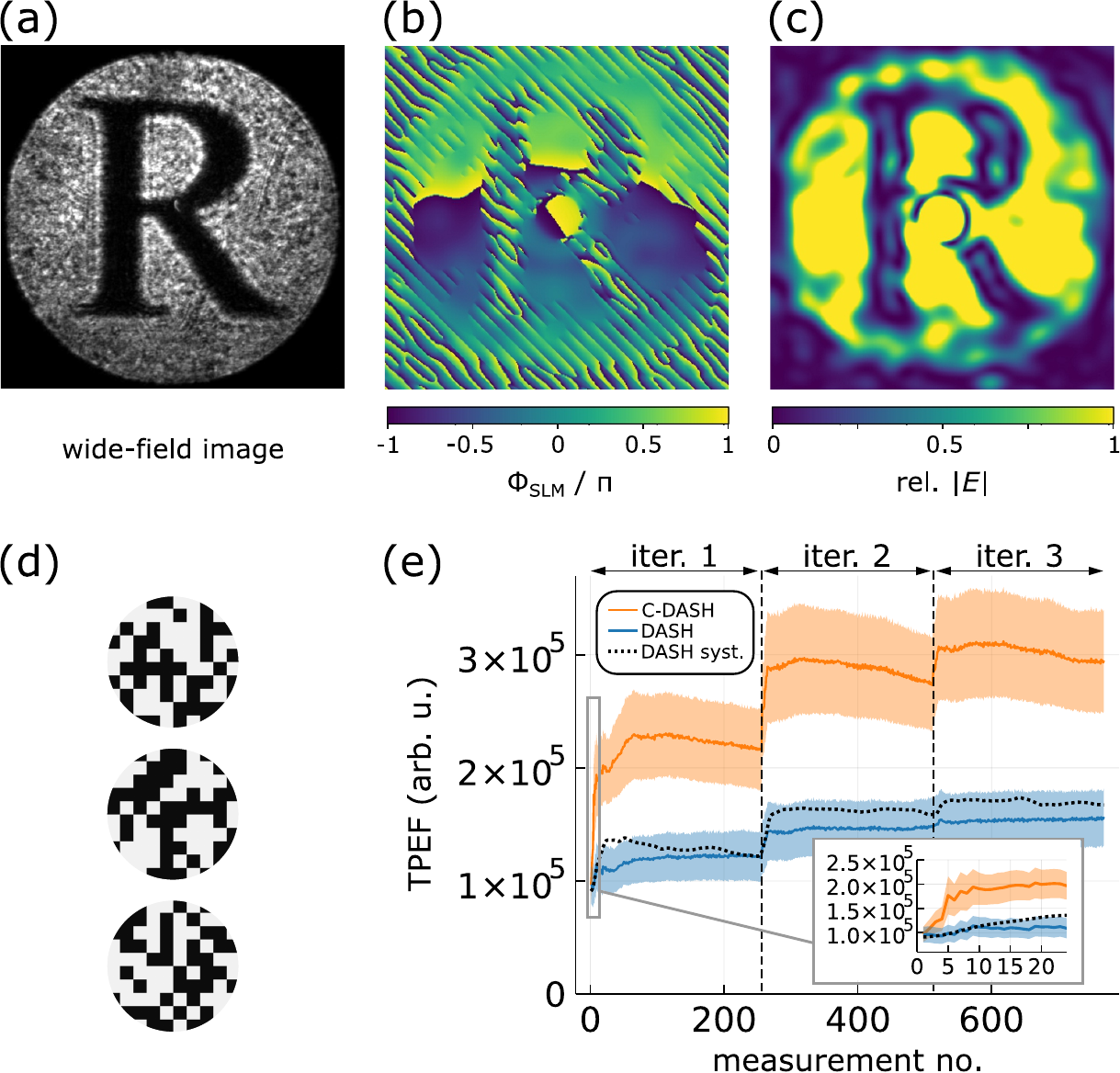}
\caption{\textbf{Experiment: avoiding absorption.} 
(a) The excitation beam is partially blocked with an ``R'' shaped mask in a pupil-conjugate plane, imaged here in wide field.
The image is black outside of the circular objective pupil. 
(b) After running the algorithm, the C-DASH mask features a high grating modulation within the ``R''-shaped obstacle and outside the objective pupil. 
(c) Light field amplitude distribution in the objective pupil produced by the phase mask in subfigure~(b). 
The ring-shaped structure in the centre is an optical system aberration (mostly spherical).
(d) Examples of 3 different, random-pixelated absorbing masks with 50\,\% filling ratio; the circle diameter corresponds to the pupil size.
(e) C-DASH (orange) vs DASH (blue) for the pixelated, absorbing scatter masks. 
The solid lines and ribbons, respectively, are the mean and standard deviation over 10 algorithm runs with different random patterns.
The black dotted line is DASH for system aberrations correction only (without absorbing mask).
The inset shows that all three lines start from the same initial TPEF level.} 
\label{fig:R-mask}
\end{figure}

Finally, we experimentally demonstrate the benefit of complex light field shaping for avoiding problematic regions in the light path.
As detailed above, `problematic' regions can be regions where the refractive index that fluctuates in time or regions that absorb light.
In this Section, we show experiments for an absorptive obstacle (which is much easier to implement in the laboratory than a fluctuating obstacle) but the results are directly transferable.
Figure~\ref{fig:R-mask}\,(a) shows an ``R''-shaped black obstacle, inserted into a pupil-conjugate plane along the excitation beam path, imaged in wide field.
After running C-DASH ($N_m=256$, $N_i=3$, $N_p=3$, $\alpha=0.5$), the correction pattern converges to the pupil-conjugate SLM mask shown in Fig.\,\ref{fig:R-mask}\,(b). 
As we see, the overlaid sawtooth grating depth is maximal at regions where light cannot reach the sample, i.e., inside the ``R''-shaped obstacle and the regions clipped by the objective pupil.
Furthermore, we see that the ``R''-shape in the phase is overlaid with a central ring structure, hinting at the presence of some systems aberrations (mostly spherical).
The effective amplitude distribution shaped by the phase pattern in Fig.\,\ref{fig:R-mask}\,(b) is plotted in Fig.\,\ref{fig:R-mask}\,(c), showing a nice agreement with the wide-field image of the ``R''-obstacle.

This example with the ``R''-obstacle also illustrates another practical advantage of C-DASH: its inherent tendency for self-alignment.
This is essentially rooted in the fact that a single device (our SLM) simultaneously carries out the two tasks of measuring the aberration modes and displaying the correction pattern.
As we see in Fig.\,\ref{fig:R-mask}\,(a--c), such a correction pattern automatically forms at the correct position, has the correct size, the correct orientation, and excitation light is redistributed from outside into the pupil.
This is a huge advantage compared with, e.g., methods such as F-SHARP, where the modes are measured using a tip-tilt piezo-actuated mirror in one location of the light path, whereas the correction pattern is displayed on an SLM in a different location of the light path.
Experimentally, this requires to find the matching relation between the two locations, necessitating an adjustment of position, size, orientation -- and sometimes numerical propagation -- of the measured correction pattern.
Such matching procedures are typically tedious, often imprecise, and may need frequent recalibration if the experimental setup is not perfectly stable regarding drifts. 

Let us turn to Fig.\,\ref{fig:R-mask}\,(d,\,e) for a more quantitative analysis.
To get statistically meaningful data, one needs to vary the absorptive obstacle for the different measurement runs (e.g., the TPEF evolution over several algorithm runs with the ``R'' mask would all look very similar).
Therefore, in this experiment, we used a set of pixelated, random absorptive patterns of 50\,\% filling ratio [examples shown in Fig.\,\ref{fig:R-mask}\,(d)].
The power parameter $\alpha$ was kept at $0.5$.
Figure~\ref{fig:R-mask}\,(d) shows the evolution of TPEF signal over the algorithm runs of C-DASH (orange) and DASH (blue), each correcting $N_m=256$~modes over $N_i=3$ iterations.
Solid lines represent the mean, shaded bands the standard deviation over 10 algorithm runs with a different absorptive pattern. 
The black, dotted line represents a DASH run with no absorptive pattern, i.e., correcting optical system aberrations only, rescaled to the same initial TPEF level.
By comparison, we see that the signal improvement observed for DASH (blue) is compatible with only the systems aberrations corrected (black, dotted line).
C-DASH, in contrast, improves the signal more than is possible by system aberrations correction only.
As we have observed before, this happens by redistribution of light from absorbing to non-absorbing regions, increasing the TPEF level by a factor of about 3 compared with the initial level. 
This is close to the theoretical limit of a factor of about $2^\ell = 4$, if we consider the power overhead of $1/\alpha = 2$ and the 50\,\% filling ratio of the absorptive pattern.

As a final remark, note that there is a strong relation between techniques that display a sequence of modes (such as C-DASH) and thus guide light around obstacles on the one hand, and the working principle of single-pixel imaging on the other hand, where unknown objects are reconstructed computationally after illuminating them with a sequence of light patterns and detecting the reflected or transmitted light on a single-bucket detector~\cite{gibson2020single}.

\section{Discussion}

Phase-only Dynamic Adaptive Scattering Compensation Holograph (DASH) -- which forms the basis for this present work -- is currently one of the fastest feedback-based scatter compensation techniques, converging faster than other techniques based on displaying a sequence of phase patterns (such as IMPACT, F-SHARP, CSA, etc.) or genetic algorithms~\cite{may2021fast}. 
This speed advantage is based on two factors: 
First, in DASH the correction pattern is continuously updated over the course of the algorithm, which leads to a faster signal increase~\cite{vellekoop2008phase, blochet_focusing_2017, may2021fast}. 
Second, it has been realised that constructing correction patterns from phase-stepped test modes (number of steps $N_p \geq 3$) only works reliably if the the excitation power ratio  between test beam $(\propto f)$ and reference beam $(\propto 1 - f)$ can be tuned such that a favourable signal-to-background ratio is ensured. 
Tuning $f$ is straightforward in DASH, but often difficult -- or impossible -- in other schemes.

Of course, a `perfect' scatter correction is only possible if the number of correction modes $N_m$ is equal to or larger than the number of scattering modes.
However, even in cases when $N_m$ is adequate, phase-only light shaping as used by DASH is only sufficient to compensate a scatterer medium if four conditions are simultaneously fulfilled: 
(i) the scatterer is a (thin) slice; (ii) the SLM is conjugate to this scattering slice; (iii) the scatterer is non-absorbing; (iv) the scatterer does not have regions where the refractive indices changes with time. 

Instead of discarding the amplitude information of the compensation pattern $C_n$ -- potentially half of the information on the scatterer -- as in the established phase-only AO approaches, in complex or C-DASH we use joint amplitude and phase shaping.
As we have shown, joint amplitude and phase shaping performs better than phase-only light shaping if any of the conditions (i--iv) is violated. 

As for other techniques, the MPEF enhancement that can be achieved using C-DASH or DASH can vary substantially depending on the characteristics of the scatterer and the fluorophore. 
Important characteristics of the scatterer include spatial frequency content and the thickness in units of the scattering mean free path, $l/l_s$~\cite{sohmen2022sensorless}. 
Important characteristics of the fluorophore include spatial distribution (localised `guide stars' vs fluorescent sea), the fluorescence strength, and bleaching behaviour~\cite{sohmen2022sensorless}.
What we have experienced over a broad variety of scatterers and samples -- in numerical simulations as well as in the laboratory -- are four main findings: 
(1) C-DASH shows a faster initial increase of signal;  
(2) C-DASH can reach a higher final signal level in the many situations where amplitude aberrations are important, including, in particular, the situations where at least one of the conditions (i--iv) is not met;
(3) C-DASH shows a robust performance which is largely insensitive to the axial placement of the correction plane (pupil- vs sample-conjugate AO); 
(4) as DASH, C-DASH is practically self-aligning, which obviates the need for tedious and often imprecise matching procedures between measured and displayed correction pattern.

As a final remark, let us highlight that the experimental demonstration of C-DASH presented in this work, using a single reflection off a phase-only SLM for joint phase and amplitude shaping, has shown clear advantages compared with phase-only light shaping \textit{despite} its obvious limitations:
{\color{\mycolour}Our proof-of-concept implementation brings experimental simplicity, but leads to drawbacks such as (a) a fraction of excitation power being wasted, (b) a finite dynamic range of  amplitude shaping, which (c) leads to local discrepancies between the displayed amplitude distribution (contained in the sawtooth supergrating) and the (ideal) target amplitude modulation.}
Using a more sophisticated experimental approach, e.g., two successive reflections off a phase-only SLM~\cite{barre_holographic_2022}, would enable joint amplitude and phase shaping while eliminating these drawbacks (a--c). 
Such an improved experimental implementation would certainly allow reaching even higher levels of experimental scatter compensation using C-DASH.
This will be a subject of further study.


\section*{Methods}

\subsection{C-DASH algorithm}

Algorithm~\ref{alg:C-DASH} outlines the C-DASH procedure used for the simulations in this publication. 
The power fraction $f$ (power in the test mode compared with total power) typically ranged between $0.1$ and $0.2$ in our simulations and experiments (no strong dependence). 
$N_\text{SLM}$ is the side length of the (square) SLM pattern in pixels, which was around $400$ (corresponding to the pupil diameter) in our experiments. 

At the start of the C-DASH algorithm, an assumption has to be made for the initial correction pattern $C$.
In past versions of our DASH algorithm~\cite{may2021fast, may2021simultaneous}, $C$ was initialised with ones. 
However, we found that in particular for weak aberrations (predominantly forward-scattering regime) the initial value can be optimized for faster convergence. 
Details on the optimal choice for the initial $C$ will be published elsewhere.  
In the first iteration, phase stepping the piston-like test mode $M_1$ is unnecessary; from the second iteration on, however, it does play a role, wherefore we include an if-statement.

\begin{figure*}
\begin{minipage}{\linewidth}
\begin{algorithm}[H]
\caption{C-DASH}
\label{alg:C-DASH}
\begin{algorithmic}[1]

\State$\mathbf{Inputs{:}}$\\
$M$, a list of $N_m$ complex-valued plane-wave input modes of size $N_\text{SLM} \times N_\text{SLM}$.\\
$U_\text{SLM}$, a real-valued phase mask of size $N_\text{SLM} \times N_\text{SLM}$.\\
$C$, a complex array of size $N_\text{SLM} \times N_\text{SLM}$, initialised with constant entries of value $c_1$.\\
$\tilde a_{n}^\text{old}$, a complex-valued vector of length $N_m$, initially filled with zeros, with the exception of $\tilde a_{1}^\text{old}$, which is set to $c_1$. \\
$G$, a real-valued $N_\text{SLM} \times N_\text{SLM}$ discretised sawtooth grating with a period of $\Lambda$ pixels.\\
$\gamma$, a normalization factor, initialised with 1. \\
$f \in [0,1]$, the fraction of total excitation light power that goes into the test modes $M_n$. \\
$\alpha \in [0,1]$, a parameter that sets the power overhead for amplitude shaping.\\
$\Call{itp}{\eta}$, a numerical interpolation function that returns the sawtooth grating depth for a given period $\Lambda$ such that the diffraction efficiency into the chosen diffraction order equals $\eta$.
\\
\Procedure{wavefront\_sensing}{$M$, $C$}
    \For{$i=1 \ldots N_i$}
    \If {$i = 1$} 
        \State $n_0 \gets 2$
    \Else
        \State $n_0 \gets 1$
        \Comment Mode $M_1$ is only relevant from 2nd iteration on
    \EndIf
    \For{$n =n_0 \ldots N_m$} 
    \State $C \gets C - \tilde a_{n}^\text{old} \, M_n$ 
    \Comment Subtract contribution of mode $n$ from last iteration
    \State $\gamma \gets \frac{1}{N_\text{SLM}} \sqrt{ \sum_{kl}{|C_{kl}|^2}}$
    \Comment Update normalisation factor
    \For{$p =1 \ldots N_p$} 
        \State $ U_1 \gets \frac{\sqrt{1-f}}{\gamma} \, C + \sqrt{f} \, M_n \exp\left(\text i\frac{2\pi}{N_p}p\right)$
        \Comment Add test mode to current correction
        \State $U_\text{SLM} \gets \Call{phi\_slm}{U_1}$
        \Comment Complex mask to holographic phase mask
        \State $P_p \gets \Call{measure\_signal}{U_\text{SLM}}$
        \Comment Read out signal for current phase step
    \EndFor
    \State $\tilde a_n \gets \frac{1}{N_p}\sum^{N_p}_{p=1}\sqrt{P_p} \, \exp(\ii\frac{2\pi}{N_p}p)$
    \Comment Compute complex-weighted power
    \State $C \gets C + \tilde a_n \, M_n$ 
    \Comment Update complex correction mask
    \State $\tilde a_{n}^\text{old} \gets \tilde a_n$
    \Comment Update contribution of mode $n$ 
    \EndFor
    \EndFor      
    \State \Return $C$
\EndProcedure

\\
\Function{phi\_slm}{$U_1$, $G$, $\alpha$}
    \State $U_2 \gets |U_1|^2$ 
    \Comment Complex amplitude to intensity
    \For{$j =1 \ldots 3$} 
        \Comment{Looping helps to stay roughly normalised  despite clamping}
        \State $U_2 \gets \alpha \, U_2/\text{mean}(U_2)$ 
        \Comment Normalise to power fraction $\alpha$
        \State $U_2 \gets \text{clamp}(U_2, 0, 1)$
        \Comment Clamp elements to range $[0,1]$
    \EndFor
    \State $\beta \gets \Call{itp}{U_2}$ 
    \Comment Diffraction efficiency to grating depth
    \State $U_3 \gets \beta \circ G $
    \Comment Element-wise multiplication with grating 
    \State $U_3 \gets \text{mod}(\text{arg}(U_1) + U_3), 2\pi)$
    \Comment Add grating and phase mask, then wrap
\State \Return $U_3$
\EndFunction

\end{algorithmic}
\end{algorithm}
\end{minipage}
\end{figure*}

\subsection{Simulation details}
\label{sec:Methods_simulations}

The voxel size of the simulation grid is $\Delta x = \Delta y = 0.3 \,\lambda_0$, $\Delta z = 1\,\lambda_0$, where $\lambda_0$ represents the vacuum wavelength.
We use numerical wave propagation as described in Ref.\,\cite{schmidt2016wave}.

\subsection{Complex-valued light shaping using a phase-only SLM}
\label{sec:Methods_ComplexShaping}

To shape a light field according to the 2D complex mask
\begin{equation}
     C(x,y) = A(x,y) \, \text{e}^{\text i \, \Phi(x,y)}
\end{equation}
using a single reflection off an SLM, we imprint the 2D phase mask $\Phi$ onto a discretised sawtooth grating $G$. 
Our sawtooth grating $G$ has a period $\Lambda \ll N_\text{SLM}$ (in pixels); in our experiments, we typically work in the diffraction order $m=0$ and choose values of $\Lambda$ around $10$.
All diffraction orders except $m$ are dumped on an iris in a Fourier plane of the SLM (see Fig.\,\ref{fig:complex_shaping_principle}).
$G$ has a \emph{locally varying} diffraction efficiency $\eta^{(m)}(x,y)$. 
For each displayed SLM mask we locally adjust $\eta^{(m)}(x,y)$ such that it produces an amplitude distribution approximately proportional to $A(x,y)$.
Of course, effectively, we can only display amplitude distributions smoothed over a lengthscale on the order of $\Lambda$.
Keeping this limitation in mind, the local diffraction efficiency $\eta^{(m)}(x,y)$ is controlled via the local peak-valley sawtooth grating depth $2\pi\,\beta(x,y)$ in radians and the parameter $0 \leq \beta \leq 1$.
The relation between $\eta^{(m)}(x,y)$ and $\beta(x,y)$ can simply be measured experimentally, or can be approximated using the analytical relationships
\begin{align}
    \eta^{(0)}[\beta, \Lambda]
    &= \frac{1}{\Lambda^2} \left|\sum_{k=1}^\Lambda \exp\left(\text i \frac{2\pi}{\Lambda} k \beta \right) \right|^2 
    \quad \text{and}
    \label{eq:eta0}\\
    \eta^{(m \neq 0)}[\beta, \Lambda]
    &= \left(\frac{\sin\left(\pi\left(m-\beta \right)\right)}{\sin\left(\frac{\pi}{\Lambda}\left(m-\beta \right)\right)}
    \cdot \frac{\sin\left(\frac{\pi m}{\Lambda}\right)}{\pi m}  \right)^2,
    \label{eq:etam}
\end{align}
cf.\ Ref.\,\cite{kress2000digital}.
Inverting the respective relation (\ref{eq:eta0} or \ref{eq:etam}), for example via numerical interpolation, allows to encode the (local) amplitude $A(x,y)$ in the (local) grating depth parameter $\beta(x,y)$ through a function we denote $B_\alpha^{(m)}[A](x,y)$:
\begin{align}
    \beta(x,y)
    &= B_\alpha^{(m)}[A](x,y) 
    \quad\text{such that}\\
    \eta^{(m)}[\beta, \Lambda](x,y) 
    &\propto A(x,y), 
\end{align}
see Alg.\,\ref{alg:C-DASH} for details.

Via the parameter $\alpha \in [0,1]$ we can deliberately set the fraction of total optical power that reaches the objective pupil, while the fraction $(1-\alpha)$ is dumped on the iris.
This power overhead allows pupil regions $(x,y)$ with high $A(x,y)$ to become brighter during the algorithm run, whereas pupil regions $(x,y)$ with low $A(x,y)$ will become darker. 
{\color{\mycolour} While the local brightness of different pupil regions changes during a C-DASH run, our clamp-and-adjust (CA) approach (lines 37--40 in Alg.\,\ref{alg:C-DASH}) returns a $\beta(x,y)$ for which the \emph{total} power in the objective pupil stays close to constant.}
The entries of the phase pattern displayed on the SLM are computed according to
{\color{\mycolour}
\begin{equation}
    \Phi_{\text{SLM}} = 
    \beta \circ G + \Phi \qquad (\tx{mod}\,2\pi),
\label{eq:PhiSLM}
\end{equation}
where $\circ$ denotes element-wise multiplication.}
For example, a sawtooth phase grating $G$ with period $\Lambda$ along dimension $x$ (in pixels) is given by
\begin{equation}
    G(x,y) = \text{arg}\left(\exp\left(\ii\frac{2\pi}{\Lambda} x \right)\right).
\label{eq:GratingG}
\end{equation}
{\color{\mycolour} Note that $\Phi_{\text{SLM}}$ wraps around $2\pi$, so we do not require a SLM stroke greater than $2\pi$ phase shift and the sawtooth grating does not restrict the shaping of the phase part of $C$ in any way.}


\section*{Funding information}
\noindent
Austrian Science Fund FWF (P32146, P36687).

\section*{Acknowledgements}
\noindent
We acknowledge valuable discussions with colleagues from our Institute.

\section*{Disclosures} 
\noindent
The authors declare no conflicts of interest.

~

\section*{Data availability} 
\noindent
Data underlying the results presented in this paper are not publicly available at this time but may be obtained from the authors upon reasonable request.


\bibliography{main}








\end{document}